\documentclass[a4paper,fleqn,usenatbib]{mnras}
\usepackage{newtxtext,newtxmath}
\usepackage[T1]{fontenc}
\usepackage{ae,aecompl}
\usepackage{color}
\usepackage{enumerate}
      
\usepackage{graphicx}	
\usepackage{amsmath}	
\usepackage{amssymb}	

\DeclareMathOperator{\sech}{sech}   
\def\msun{\,M_\odot}
\def\pc{\,{\rm pc}}
\def\kpc{\,{\rm kpc}}
\def\kms{\,{\rm km\,s^{-1}}}
\def\myr{\,{\rm Myr}}
\def\gyr{\,{\rm Gyr}}
\def\feh{\left[{\rm Fe/H}\right]}

\def\afe{\left[{\rm \alpha/Fe}\right]}
\def\V6{V$\alpha$9s8$\lambda\zeta$*}
\def\hf{h_R(R_0,t_{\rm f})}

\title[Thin and thick disc migration and kinematics]{Migration and kinematics in growing disc galaxies with thin and thick discs}
\author[M. Aumer, J. Binney and R. Sch{\"o}nrich]
{Michael Aumer \thanks{E-mail:Michael.Aumer@physics.ox.ac.uk (MA)}, 
James Binney and Ralph Sch{\"o}nrich\\
Rudolf Peierls Centre for Theoretical Physics, 1 Keble Road, Oxford, OX1 3NP, UK}

\date{Accepted 2017 June 12. Received 2017 June 12; in original form 2017 April 25}
\pubyear{2017}
\begin{document}
\label{firstpage}
\pagerange{\pageref{firstpage}--\pageref{lastpage}} 
\maketitle

\begin{abstract}
We analyse disc heating and radial migration in $N$-body models of growing
disc galaxies with thick and thin discs. Similar to thin-disc-only models,
galaxies with appropriate non-axisymmetric structures reproduce observational
constraints on radial disc heating in and migration to the Solar
Neighbourhood (Snhd). The presence of thick discs can suppress
non-axisymmetries and thus higher baryonic-to-dark matter fractions are
required than in models that only have a thin disc. Models that are baryon-dominated to
roughly the Solar radius $R_0$ are favoured, in agreement with data for the
Milky Way. For inside-out growing discs, today's thick-disc stars at $R_0$
are dominated by outwards migrators. Whether outwards migrators are
vertically hotter than non-migrators depends on the radial gradient of the
thick disc vertical velocity dispersion. There is an effective upper boundary
in angular momentum that thick disc stars born in the centre of a galaxy can
reach by migration, which explains the fading of the high $\afe$
sequence outside $R_0$. Our models compare well to Snhd
kinematics from RAVE-TGAS. For such comparisons it is important to take into
account the azimuthal variation of kinematics at $R \sim R_0$ and biases from
survey selection functions.  The vertical heating of thin disc stars by giant
molecular clouds is only mildly affected by the presence of thick discs. 
Our models predict higher vertical velocity dispersions for the oldest stars
than found in the Snhd age velocity dispersion relation, possibly because
of measurement uncertainties or an underestimation of the number of old 
cold stars in our models.
\end{abstract}


\section{Introduction}

\citet{gilmore} discovered that the vertical star count density in the
Milky Way (MW) near the Sun could be fitted by a double-exponential profile. The two
components have become known as the {\it thin disc} and the {\it thick disc}.
\citet{juric} find scaleheights of $h_{z, {\rm thin}}=300\pc$ and $h_{z, {\rm
thick}}=900\pc$ and a local thick disc contribution to the stellar surface
density of $\sim 26$ per cent.  The majority of edge-on disc galaxies in the
local Universe prove to have   vertical surface brightness profiles that are
consistent with a similar double-exponential structure
\citep{comeron}. 

The thick part of the density profile is made up of stellar populations with
hotter vertical kinematics, which in comparison to the thin populations, have
a different chemical composition in the form of enhanced $\alpha$-element
abundances relative to their iron content, $\afe$ \citep{fuhrmann}, and older
ages \citep{masseron}. High $\afe$ populations have been found to be more
centrally concentrated than thin disc stars \citep{bensby, cheng} and to fade
outside a Galactic radius of $R\sim10\kpc$ \citep{hayden}.  Radial age
gradients at high altitudes $|z|>1\kpc$ \citep{martig16} suggest that
chemical and geometrical definitions of the thick disc can yield very
different results in the outer Galaxy.  The formation scenarios for thick
discs split mainly into two groups: (i) the heating of an initially thin disc
in galaxy mergers (e.g. \citealp{quinn}), or (ii) the birth of thick disc
stars on kinematically hot orbits (e.g. \citealp{bird}). For (ii) both the
formation in an early gas-rich and turbulent phase dominated by mergers and
high gas accretion rates \citep{brook} and the continuous decrease of birth
velocity dispersions of stars in discs with declining gas fractions and
turbulence driven by gravitational disc instabilities \citep{forbes} have
been suggested. 

The diverse chemical compositions of stars in the Solar Neighbourhood (Snhd)
favour models in which Snhd stars were born at a variety of Galactic radii.
The chemical pattern in thick-disc stars requires that these stars originated
in significantly more central regions than the Snhd \citep{sb09a} and that
the Galaxy has undergone an inside-out formation scenario \citep{ralph2017}.
The high $\alpha$ abundances at high iron abundance $\feh$ of some thick-disc
stars require very high star formation efficiencies in the early MW
\citep{brett} and thus a compact star-forming disc.

The process that has been identified as responsible for spreading stars away from their birth radii is 
radial migration. It happens when stars are scattered by a non-axisymmetric structure such as a bar or 
a spiral arm across the structure's corotation resonance, which leads to their angular momentum changing 
appreciably without significantly increasing random motions \citep{sellwoodb}. It can explain the 
chemical diversity of Snhd stars \citep{sb09a,sb09b} and the dependence of the specific shape of stellar 
metallicity distribution on Galactic radius \citep{loebman16}.

\citet{sb09b} and \citet{ralph2017} showed that in their analytical Galaxy
models, which couple dynamics and chemical evolution, radial migration is
responsible for the formation of the thick disc as outwards migrating stars
from hot inner disc parts provide a population of vertically hot stars at
outer radii.  The source of vertical heating in the inner disc was left
unspecified in this work, which assumed that the velocity dispersions of a
stellar population are specified by its birth radius and age through a
combination of: (i) an inwards increasing vertical velocity dispersion, that
would ensure a radially constant scaleheight in the absence of radial
migration, and (ii) a radially independent time dependence of the vertical
heating law $\sigma_z(t)$. Whereas \citet{sb09b} assumed that during
migration the energy of vertical oscillations is conserved, \citet{ralph2017}
corrected this assumption to conservation of vertical action \citep{solway}.

Simulations of disc galaxies forming from the cooling of rotating hot gas
haloes have roughly confirmed this picture \citep{loebman, roskar} and
studies of radial migration in thick and thin disc systems found mildly lower
but significant levels of migration in the thick populations \citep{solway}.
However, several other studies have cast doubt on the migration origin of the
thick disc \citep{minchev12, vera14,vera16}, as they concluded that
vertically hot stars are less prone to radial migration than vertically cool
stars.

Important observational constraints on the evolution of the MW come from the age
velocity dispersion relation (AVR), which shows that bluer and therefore younger
populations have lower velocity dispersions in all three directions $R$, $\phi$ and 
$z$ than  redder, older  populations \citep{parenago, wielen,db98,holmberg}. 
Interestingly, if one uses stars close to the Sun from the Geneva Copenhagen Survey
(GCS, \citealp{nordstrom, casagrande11}), the oldest stars show vertical velocity 
dispersions $\sigma_z\sim30\kms$. This is significantly lower than the 
$\sigma_z\sim40-50\kms$ found for high $\afe$ stars by \citet{bovy}. Moreover, the
vertical AVR shows no step in time, as might be expected from a merger scenario, and
as had been claimed to have been found in the Snhd \citep{quillen}.

In \citet{abs16a} (hereafter Paper 1), we presented a set of $\sim100$
$N$-body simulations of disc galaxies growing within live dark matter (DM)
haloes over $\sim10\gyr$. In \citet{abs16b} (hereafter Paper 2), we analysed
the stellar kinematics of these models to get a better understanding on disc
heating in MW-like disc galaxies. Papers 1 and 2 established that giant
molecular clouds (GMCs) are a necessary ingredient for reproducing disc
structure and kinematics as they are the main vertical heating agent for the
thin disc and can explain the vertical thin-disc AVR in the Snhd. Moreover,
we showed that models which combined the standard values for dark-halo mass and
baryonic Galaxy mass could reproduce the radial AVR and the amount of
migration required to explain the chemistry of the Snhd. Because both quantities are
determined by non-axisymmetric disc structures, we can assume that they were
appropriately captured by the models.

As none of these models contains a thick disc similar to that in the MW,
\citet{ab17} (hereafter Paper 3) introduced a new set of simulations in which
thick discs were either present in the initial conditions (ICs) or created by
adding stars on orbits with continuously decreasing velocity dispersions.
Paper 3 showed that several of these models are capable of simultaneously
roughly reproducing for the MW the vertical and radial mass distributions, the
length and edge-on profile of the bar, vertical and radial age gradients of disc stars
and constraints on the  DM density in the Snhd and the circular-speed
curve.

Here we use a subset of the simulations from Papers 1 and 3 to study radial
migration and Snhd kinematics in models with thin and thick discs. We clarify
the extent to which the presence of a thick disc modifies our conclusions
regarding disc heating and migration, and we quantify migration of thick-disc
stars. We use the models to investigate the impact of the Galactic bar and
spiral structure on measurements of disc kinematics such as those recently
released by the Gaia consortium \citep{tgas}.

The paper is structured as follows. Section \ref{sec:simulations} summarises
the setup and parameters of the simulations and can be skipped by a reader
familiar with Paper 3. Section \ref{sec:mig} quantifies the radial migration by
stars of various ages. Section \ref{sec:kinematics} compares the stellar
kinematics extracted from our models with data from the MW: Section
\ref{sec:tgas} analyses velocity distributions in the $R$, $\phi$, and $z$
directions and compares them to recent data from RAVE and TGAS; Section
\ref{sec:agevel} discusses how well AVRs of models with thin and thick discs
compare to Snhd data. In Section~\ref{sec:discuss} we discuss our results with an
emphasis on the role of radial migration in the formation of the 
chemically-defined thick disc (Section \ref{sec:di:mi}) and on how disc heating 
shapes Snhd kinematics (Section \ref{sec:di:ki}). Section \ref{sec:conclude} sums up.

\tabcolsep=2.5pt
\begin{table*}
\vspace{-0cm}
  \caption{List of models analysed in this paper and their parameters. 
           {\it 1st Column}: Model Name;
           {\it 2nd Column}: Initial Conditions;
           {\it 3rd Column}: Initial baryonic mass $M_{\rm b,i}$;
           {\it 4th Column}: Concentration parameter for IC dark halo $c_{\rm halo}$;
           {\it 5th Column}: IC scalelength $h_{\rm IC}$;
           {\it 6th Column}: IC disc scaleheight $z_{0, {\rm disc}}$;
           {\it 7th Column}: Total inserted baryonic model mass $M_{\rm f}$ 
           (including initial baryonic mass);
           {\it 8th Column}: Final time $t_{\rm f}$;   
           {\it 9th Column}: Initial disc scalelength $h_{R, {\rm i}}$;
           {\it 10th Column}: Final disc scalelength $h_{R, {\rm f}}$;
           {\it 11th Column}: Scalelength growth parameter $\xi$;
           {\it 12th Column}: Stellar exponential scalelength $\hf$ determined from all stars within $5.8<R/\kpc<10.8$ at $t_{\rm f}$ ;
           {\it 13th Column}: Type of SFR law;
           {\it 14th Column}: Exponential decay timescale $t_{\rm SFR}$ for the star formation rate;
           {\it 15th Column}: Radial to vertical dispersion ratio for inserted particles $\lambda$;
           {\it 16th Column}: Prescription for initial velocity dispersion for inserted stellar particles, {$\sigma_0(t)$};
           {\it 17th-18th Column}: Parameters $\sigma_1$ and $t_1$, which determine {$\sigma_0(t)$};
           {\it 19th Column}: GMC star formation efficiency $\zeta$.}
  \begin{tabular}{@{}ccccccccccccccccccc@{}}\hline
1st & 2nd &  3rd      & 4th     &  5th     &  6th          & 7th       & 8th      & 9th        &  10th  &    11th& 12th &13th   & 14th        & 15th      & 16th      & 17th       & 18th & 19th    \\
Name&ICs&$M_{\rm b,i}$&$c_{\rm halo}$&$h_{\rm IC}$&$z_{0,{\rm disc}}$&$M_{\rm f}$&$t_{\rm f}$&$h_{R,{\rm i}}$&$h_{R,{\rm f}}$&$\xi$&$\hf$&SF Type&{$t_{\rm SFR}$}&{$\lambda$}&{$\sigma_0$}&{$\sigma_1$}&$t_1$ &{$\zeta$}\\ 
      &  &{$[10^{9}\msun]$}&   &{$[\kpc]$}&{$[\kpc]$}&{$[10^{10}\msun]$}&{$[\gyr]$}&{$[\kpc]$}&{$[\kpc]$}     &  & $[\kpc]$&    &{$[\gyr]$}    &           & Type      &{$[\kms]$}&{$[\gyr]$}& \\ \hline
   
Y1    & Y   &  5      & 9.0     & 1.5        & 0.10          & 5        & 10       & 1.5        & 4.3       & 0.5 & 4.0 & 1       & 8.0         & 1.0       & const      & 6         & --   &  0.08   \\ 
\hline
P1s6  & P   &  15     & 9.0     & 2.5        & 1.75          & 5        & 10       & 1.5        & 4.3       & 0.5 & 2.8 & 0       & --          & 1.0       & const      & 6         & --   &  0.08   \\
P2    & P   &  15     & 9.0     & 2.5        & 1.75          & 5        & 10       & 2.5        & 2.5       & 0.0 & 3.0 & 1       & 8.0         & 1.0       & const      & 6         & --   &  0.08   \\
Q1$\zeta$-& Q& 25     & 6.5     & 2.5        & 1.75          & 6        & 10       & 1.5        & 4.3       & 0.5 & 4.1 & 1       & 8.0         & 1.0       & const      & 6         & --   &  0.04   \\
U1    & U   &  25     & 7.5     & 2.0        & 1.70          & 6        & 10       & 1.5        & 4.3       & 0.5 & 5.8 & 1       & 8.0         & 1.0       & const      & 6         & --   &  0.08   \\
\hline\hline
M$\alpha$1& M&  5     & 9.0     & 1.5        & --            & 5        & 10       & 1.5        & 4.3       & 0.5 & 3.2 & 1       & 8.0         & 1.0       & atan       & 16        & 2.0  &  0.08   \\
M$\beta$1s5& M&  5    & 9.0     & 1.5        & --            & 5        & 12       & 1.5        & 4.3       & 0.5 & 2.8 & 1       & 8.0         & 1.0       & plaw       & 51        & 1.57 &  0.08   \\
V$\alpha$8s7& V& 5    & 6.5     & 1.5        & --            & 6        & 12       & 1.0        & 4.3       & 0.6 & 3.7 & 2       & 12.0        & 1.0       & atan       & 16        & 3.5  &  0.08   \\
V$\beta$8s5& V&  5    & 6.5     & 1.5        & --            & 6        & 12       & 1.0        & 4.3       & 0.6 & 5.6 & 1       & 8.0         & 1.0       & plaw       & 51        & 1.57 &  0.08   \\
V$\alpha$9s8$\lambda\zeta$*& V& 5 & 6.5 & 1.5 & --           & 6        & 12       & 1.0        & 3.5       & 0.6 & 3.1 & 1       & 6.0         & 1.25      & atan       & 16        & 2.0  &  0.06   \\
\hline
  \end{tabular}
  \label{modeltable}
\end{table*}

\section{Simulations}
\label{sec:simulations}

Table~\ref{modeltable} lists the small subset of the simulations presented in
Papers 1 and 3 that are discussed in this paper. We discuss only collisionless
simulations of growing disc galaxies within non-growing live dark haloes made using
the Tree code \textsc{gadget-3}, last described in \citet{gadget2}.  We distinguish between 
(i) thin-disc-only simulations (Y1), (ii) models with thick discs in their ICs 
(P, Q and U models), and (iii) models in which stars are added with continuously 
declining velocity dispersions (declining $\sigma_0$ models with V and M ICs).

A model is specified by the initial conditions from which it starts and the
rules used to feed in stars.  For a full account of the meaning of all
parameters and the model names we refer to Papers 1 and 3. Here we only give
a brief overview. In the Appendix \ref{appendix}, we present figures which
show basic structural properties for each of the models studied in this paper.

We focus on standard-resolution models, which at the end of the simulation
contain $N_{\rm b}=5-6\times10^6$ stellar particles with particle masses
$m_{\rm b}=10^{4}\msun$ and $N_{\rm DM}=5\times10^6$ DM
particles with masses $m_{\rm DM}=2\times10^{5}\msun$. In
addition, the simulations contain a population of short-lived, massive
particles representing GMCs with masses $m_{\rm GMC}=10^{5-7}\msun$ following
a GMC mass function similar to that observed in the MW. The force softening
lengths are $\epsilon_{\rm b}=30\pc$ for baryonic particles (including GMCs)
and $\epsilon_{\rm DM}=134\pc$ for DM particles.

\subsection{Initial conditions}

Table 1 of Paper 3 lists the details of the ICs, which were created using the \textsc{galic}
code \citep{yurin}. All models discussed here start with a spherical dark halo with a 
\citet{hernquist} profile
\begin{equation}
\rho_{\rm{DM}}(r)={{M_{\rm{DM}}}\over{2\pi}} {{a}\over{r\left(r+a\right)^3}}.
\end{equation}
and mass $M_{\rm DM}=10^{12}\msun$. The inner density profiles are adjusted
to be similar to NFW profiles with concentration parameters in the range
$c_{\rm halo}=6-9$.

The ICs contain either a stellar disc (Y, P, Q or U ICs) or a low-density
distorted stellar bulge resembling a rotating elliptical galaxy  (M
and V ICs). IC discs have a mass profile
\begin{equation}
{\rho_{\rm{disc,i}}(R,z)} = {{M_{\rm{b,i}}}\over{4\pi {z_{0,{\rm
disc}}}{h_{\rm IC}}^2}} {\sech^2 \left({z}\over{z_{0,{\rm
disc}}}\right)} {\exp\left(-{R}/{h_{\rm IC}}\right)}.
\end{equation}
Here $h_{\rm IC}$ is the IC disc exponential scalelength and
a radially constant isothermal vertical profile with scaleheight $z_{0, {\rm disc}}$
is assumed. The Y model starts with a baryonic disc of mass $M_{\rm b,i}=5\times10^9\msun$,
which is compact ($h_{\rm IC}=1.5\kpc$) and thin ($z_{0, {\rm disc}}=0.1\kpc$), whereas
the P, Q and U models contain a much thicker, more extended and more massive IC disc 
($z_{0, {\rm disc}}\sim1.7\kpc$, $h_{\rm IC}=2.0-2.5\kpc$, $M_{\rm b,i}=15-25\times10^9\msun$).

The baryonic ICs of the M and V models have a Hernquist profile with $M_{\rm b,i}=5\times10^9\msun$
and $a=h_{\rm IC}=1.5\kpc$. The mass profile is distorted to an oblate spheroid with axis ratio
$s=2$ as $\rho_{\rm elliptical}(R,z)=s\rho_{\rm Hernquist}\Bigl(\sqrt{R^2+s^2z^2}\Bigr)$.

\subsection{Growing the discs}

Stellar particles are continuously added to the disc following a star formation rate ${\rm SFR}(t)$
that is either constant in time (type 0), declines exponentially as
\begin{equation}\label{eq:SFTt}
{\rm SFR}(t)={\rm SFR}_0 \times \exp({-t/t_{\rm SFR}})\,\,\, {\rm {(type \,1)}},
\end{equation}
or has an additional early increase
\begin{equation}
{\rm SFR}(t)={\rm SFR}_0 \times \exp({-t/t_{\rm SFR}}-{0.5\gyr/t})\,\,\, {\rm {(type \,2)}},
\end{equation}
with $t_{\rm SFR}=6-12\gyr$. The constant ${\rm SFR}_0$ is adjusted to
produce at $t=t_{\rm f}$ a target final baryonic mass $M_{\rm f}$ in the range
$5-6\times10^{10}\msun$ including the IC mass $M_{\rm b,i}$.

As was discussed in Papers 1 and 3, mass growth and the influence of bars and spirals
change the surface density profiles of stars of all ages and lead to final profiles which differ
from simple exponentials. To characterise the total final stellar surface density profile 
$\Sigma(R)$ in a Snhd-like location, we fit an exponential to $\Sigma(R)$ at $t_{\rm f}$ in the region 
$5.8<R/\kpc<10.8$ centred around $R_0=8.3\kpc$ and give the resulting scalelength $\hf$ in Table 1. We note that not
all of the profiles are well-fitted by exponentials in this radial range.

Every five Myr, stellar particles are added at $z=0$ and randomly chosen azimuths, and with 
an exponential radial density profile $\Sigma_{\rm SF}(R)\propto\exp(-R/h_R(t))$. The 
scalelength $h_R(t)$ of the newly added particles grows in time as
\begin{equation}\label{eq:hRt}
h_R(t)=h_{R,\rm i}+(h_{R,\rm f}-h_{R,\rm i})(t/t_{\rm f})^\xi.
\end{equation}
To avoid inserting particles in the bar region, particles are not added inside a cutoff 
radius $R_{\rm cut}$, which is determined by the current bar length as measured by the $m=2$ 
Fourier amplitude $A_2(R)$. If there is a peak in $A_2$ with $\ln(A_2)>-1.5$ in the inner 
galaxy, we set $R_{\rm cut}$ to the smaller of 5 kpc and the radius where $\ln(A_2)(R)$ drops 
below $-1.5$. Our choice for the cutoff is empirical in nature and was discussed
in Paper 1. We find that the upper limit for the cutoff length does not prevent longer bars 
from forming (e.g. in model U1) and that these longer bars are not significantly different 
from shorter bars. We find that our cutoff definition is reasonably well suited for the models 
presented here as it picks out the strong bar regions and avoids contributions from spirals (see 
e.g. Figure \ref{sd}). We caution that this need not be the case for any possible model.

GMCs are modelled as a population of massive collisionless particles drawn from a mass
function of the form ${\rm d}N/{\rm d}M\propto M^\gamma$ with lower and upper mass 
limits $M_{\rm low}=10^5\msun$ and $M_{\rm up}=10^7\msun$ and an exponent $\gamma=-1.6$.
GMC particles are added in the same region, with the same profile as stellar particles and on orbits
with birth velocity dispersion $\sigma_0=6\kms$ (see Section \ref{sec:bvd}). Their azimuthal density is given by
\begin{equation}
\rho_{\rm GMC}(\phi)\propto \left[\rho_{\rm ys}(\phi)\right]^\alpha,
\end{equation}
where $\rho_{\rm ys}$ is the density of young stars and $\alpha=1$. The mass
in GMCs is determined by the SF efficiency $\zeta=0.04-0.08$.  Specifically, for each
$\Delta m_{\rm stars}$ of stars formed, a total GMC mass $\Delta m_{\rm GMC}
= \Delta m_{\rm stars}/\zeta$ is created. GMC particles live for only
$50\myr$: for $25\myr$ their masses grow with time as $m\propto t^2$, and for the
final $25\myr$ of their lives their masses are constant. At early times, GMCs
can contain a substantial fraction of the baryonic galaxy mass, especially
for models with low IC masses, low $\zeta$ and short $t_{\rm SFR}$. For model
\V6 at early times, 45 per cent of the baryonic mass is in GMCs. At late times, GMC
mass fractions are low: $2-3$ per cent.

\begin{figure}
\vspace{-0.cm}
\includegraphics[width=8cm]{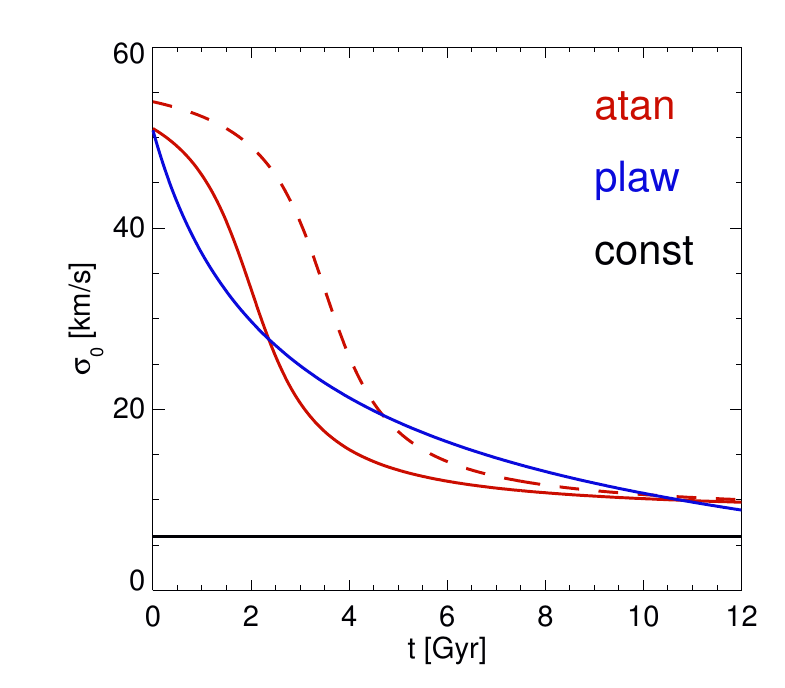}\\
\caption
{Evolution of input velocity parameter $\sigma_0$ with simulation time $t$ for various scenarios .
The black line is for type `const', the red lines for type `atan' (dashed for $t_1=3.5\gyr$, solid for
$t_1=2.0\gyr$) and the blue line for type `plaw'.}
\label{sig0}
\end{figure}

\subsection{Birth velocity dispersions}
\label{sec:bvd}

The young stellar populations are assigned birth velocity dispersions
$\sigma_0(t)$.  For thin-disc-only simulations and models with thick-disc ICs
we assign a constant $\sigma_z=\sigma_R=\sigma_{\phi}=\sigma_1=6\kms$. The
mean rotation velocity $\langle v_{\phi}\rangle(R)$ at radius $R$ is set to the circular
speed $v_{\rm circ}=\sqrt{{a_R(R)} R}$, where $a_R(R)$ is the azimuthal
average of the radial gravitational acceleration, $\partial\Phi/\partial R$.

For declining $\sigma_0$ models we use two different functional forms for $\sigma_0(t)$:
\begin{equation}\label{atan}
\sigma_0(t)=\sigma_1 \left[ {\arctan \left( {t_1-t} \over {1\gyr}  \right)} + {{\pi} \over {2}} \right]+ 
8\kms \,\,\, {\rm (type \, atan)}
\end{equation}
or
\begin{equation}\label{plaw}
\sigma_0(t)=\sigma_1 \left( {t+t_1} \over {2.7\gyr} \right)^{-0.47} -15\kms \,\,\, {\rm (type \, plaw)}.
\end{equation}
We assume that $\sigma_0$ is independent of radius $R$ and always set
$\sigma_z=\sigma_0$ and $\sigma_{\phi}=\sigma_R/\sqrt{2}$. We test values
$\sigma_R=\lambda\sigma_0$ with $\lambda=1-1.3$ and apply an asymmetric drift
correction for $\langle v_{\phi}\rangle(R)$ derived from Equation (4.228) in
\citet{gd2}.  Figure \ref{sig0} visualises the difference between assumed
histories of the decline in $\sigma_0$.  For models with declining $\sigma_0$
we choose a higher thin-disc input dispersion $\sigma_0\sim10\kms$ than in
the models with constant $\sigma_0=6\kms$. As was shown in Paper 2 this
yields a somewhat better fit to Snhd AVRs. The AVRs of the present
simulations are discussed in Section \ref{sec:agevel}.

\begin{figure*}
\vspace{-0.cm}
\hspace{-0.1cm}\includegraphics[width=18cm]{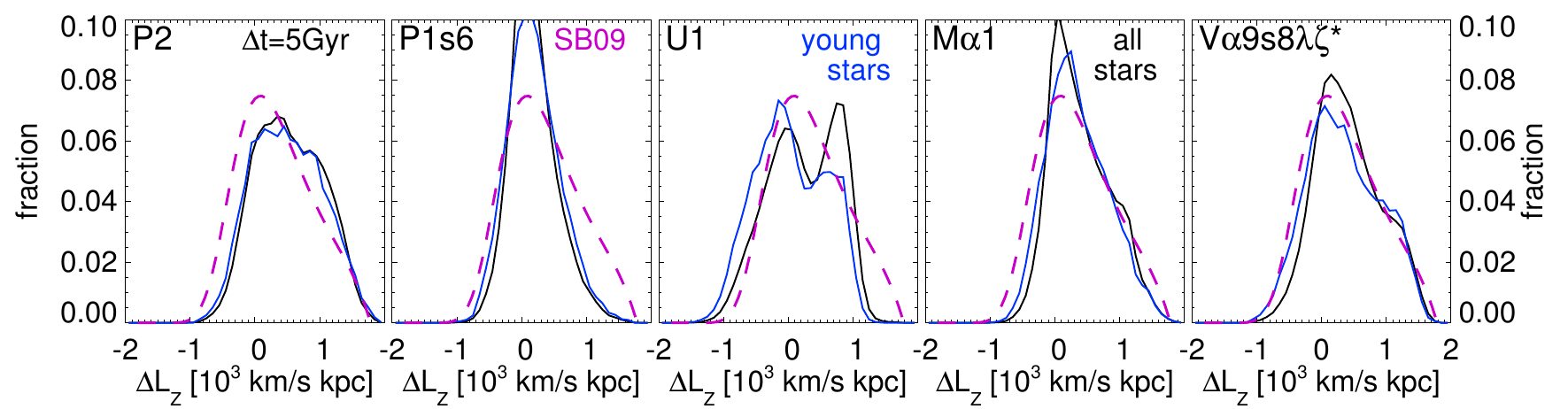}\\
\caption
{Plots of changes in $L_z$ in the last $5\gyr$ for samples of stars selected to have $L_z=L_{z,{\rm circ}}(8\kpc)\pm100\kpc\kms$ 
at $t=t_{\rm f}$ in various models. The black curves are for all stars that are more than $5\gyr$ old, while the blue
curves are for stars with ages $5-6\gyr$, so $\Delta L_z$ is the change in their angular momentum since they were young.  
Also plotted in a broken line is the corresponding distribution inferred by \citet{sb09a} from the chemical
composition of Snhd.}
\label{migthin}
\end{figure*}

\section{Radial Migration}
\label{sec:mig}

We begin the analysis of the selected models by studying radial migration of
disc stars of various ages and over different timescales to gain a better
understanding of how the models differ and to understand how well they fulfil
constraints set by chemical evolution models, which require a significant
level of radial migration to explain the variety of stellar chemical
compositions found in the Snhd \citep{sb09a}.

\subsection{Thin-disc constraints}
\label{sec:migthin}

With Figure \ref{migthin} we test whether the conclusion from Paper 1, that
thin-disc-only models of appropriate mass in appropriate dark haloes provide
the right amount of migration to explain the chemistry of the Snhd is also true for
thin disc components in thin+thick disc systems.  We plot histograms of the
change $\Delta L_z\equiv L_z(t_{\rm f})-L_z(t_{\rm f}-5\gyr)$ in the angular
momentum of stars which at $t=t_{\rm f}$ have angular momentum in the range
$L_z=L_{z,{\rm circ}}(8\kpc)\pm100\kpc\kms$.  Here $L_{z,{\rm circ}}\equiv R
v_{\rm circ}(R)$ and these stars thus have current guiding centre radii
$R_{\rm g}\approx8\kpc$. The black curves include all stars born before
$t_{\rm f}-5\gyr$, while the blue curves are for stars younger than $1\gyr$
at that time, so $\Delta L_z$ is the change in their angular momentum since
they were young. The pink dashed curve is from the chemical evolution model
of \citet{sb09a}. 

Model P2 shows, both for all stars, and just young stars, a $\Delta L_z$
histogram which is very similar to the one predicted for the Snhd by
\citet{sb09a}. It is also similar to the histogram of Model YG1 presented in Paper 1, which
demonstrates that the presence of a thick disc and the relatively late
formation of a bar in P2 do not suppress migration to $R=8\kpc$.  Model P1s6
has less non-axisymmetric structure than P2 (see Paper 3 and Figure \ref{sd}), 
so its $\Delta L_z$ distribution is significantly narrower than required by Snhd chemistry.

The histogram for Model U1 shows a sharper cutoff at high $\Delta L_z$.
As was discussed in Paper 1, this cutoff is caused by the early presence of a
relatively long bar, out of which stars cannot escape to reach radii close to
$R_0$.  Curiously, especially the old stars show a double-peaked $\Delta L_z$
histogram. This feature arises already between $t=5$ and $6\gyr$, when it is
more strongly detectable and is connected to the corotation resonance of the
bar which forms at $t\sim5\gyr$ (see Figure \ref{barx}).

Models M$\alpha$1 and \V6 have declining $\sigma_0$. The choice of a radially 
constant input dispersion $\sigma_0$ leads to on average higher velocity dispersions at 
outer radii than in models with thick-disc ICs. This in turn reduces the 
strength of non-axisymmetric structures, as shown in Figure \ref{sd}. 
Consequently, Model M$\alpha$1, despite having
a reasonably sized bar, shows mildly too little radial migration. By contrast, \V6,
which has a more massive and more compact disc and lives in a lower density
halo, has a distribution of $\Delta L_z$ that comes very close to that
inferred from observational data by \citet{sb09a}.

Note that the difference between all stars and young stars is not significant
for any models. Given that in all models young populations are on average
vertically cooler than the average over all stars, which includes the thick
disc, this implies significant migration taking place for vertically
hotter stars. We will examine this further below.

In light of these results, it is interesting to reconsider the results of Figure 9 
of Paper 3. This figure plots the evolution of the exponential scalelengths of
mono-age populations measured at a Solar-like radius $h_R(R_0)$. It shows that any model 
that features the required amount of radial migration also shows an increase 
during the last $5\gyr$ in $h_R(R_0)$ for populations of all ages. So, the 
non-axisymmetries in the MW that caused radial migration likely also
affected the radial surface density profiles of stars at $R_0$ in a flattening way.

\subsection{Migration to the Snhd as a function of age}
\label{sec:migsnhd}

\begin{figure*}
\hspace{-0.3cm}\includegraphics[width=18cm]{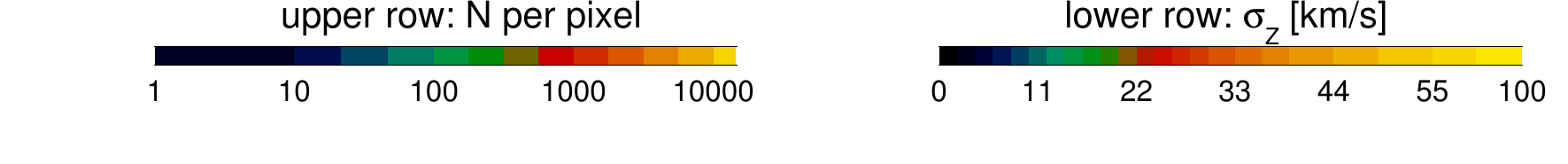}\vspace{-0.7cm}\\
\hspace{-0.3cm}\includegraphics[width=18cm]{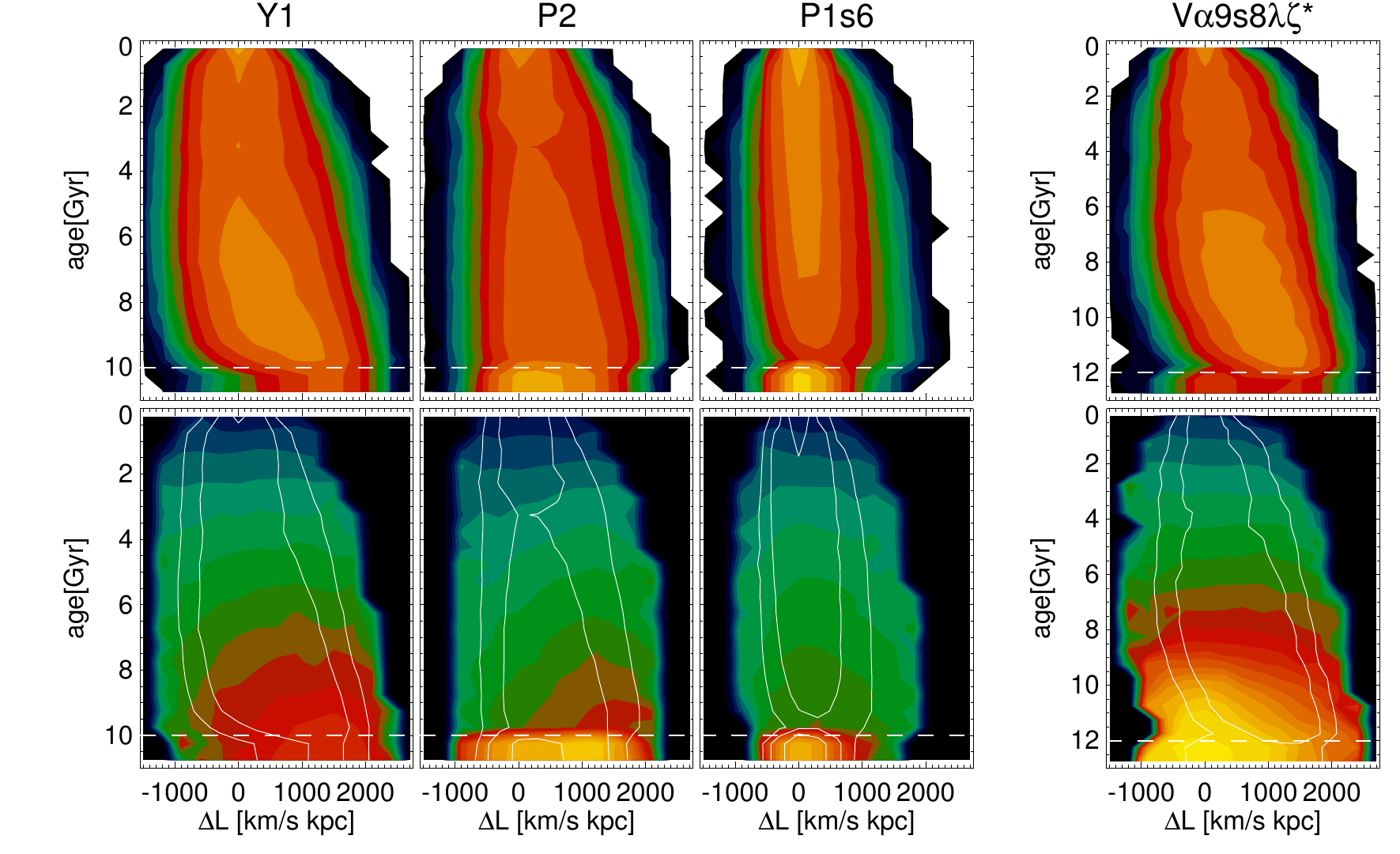}\\
\caption {Upper panels: the density of stars, which at $t=t_{\rm f}$ are at
$R=8\pm0.5\kpc$, in the $\tau$-$\Delta L_z$ plane.  $\tau$ is the age of a
star and $\Delta L_z= L_z(t_{\rm f})-L_z(t_{\rm birth})$ is the change in
angular momentum a star has undergone since birth. The horizontal dashed
lines mark the distinction between IC stars and added stars.  Lower panels: the
vertical velocity dispersion $\sigma_z$ at $t=t_{\rm f}$ of stars at
$R=8\pm0.5\kpc$ in the $\tau$-$\Delta L_z$ plane.  White contours are for
density as in the upper row.} \label{migall}
\end{figure*}

In Figure \ref{migall} we analyse the origin of all stars that at $t=t_{\rm f}$ live 
at $R=8\pm0.5\kpc$. Note that unlike in Section \ref{sec:migthin} we here select stars
in $R$ and not $L_z$. The upper row of panels shows the distribution of these stars in
the plane of age $\tau$ and $\Delta L_z= L_z(t_{\rm f})-L_z(t_{\rm birth})$, the angular 
momentum change since birth. The horizontal dashed lines mark the distinction between
IC stars and added stars.

We start by analysing the inside-out growing thin-disc-only model Y1. We
notice that the youngest stars, which have not had enough time to
migrate significantly, show a rather even distribution around $\Delta L_z=0$.
Towards the older added and IC stars the distribution bends clearly towards
positive $\Delta L_z$, i.e. the majority of stars were born at smaller $L_z$
and have migrated outwards. Y1 has a compact early disc with $h_R=1.5\kpc$
both for IC stars and the first inserted stars, which means there simply aren't
many stars at $R=8\kpc$ at these times. Moreover, non-axisymmetric structures
are strong from relatively early on (Figure \ref{barx}) and the populations are
all initially cold, which leads to significant migration of these old
populations over the course of the simulations.

Model P2 has a thick, more extended IC disc with $h_R=2.5\kpc$ and an
insertion scalelength that is constant at the same value. We see two main
differences. (i) Unlike in Y1, there is a discontinuity in the $\Delta L_z$
distribution between the IC stars and the old inserted stars. This is because
IC stars have a different original $L_z$ distribution and are vertically
hotter than the other populations, which are all born cold and,
notwithstanding GMC heating, have at all times lower $\sigma_z$ than the IC
stars (see Section \ref{sec:agevel}). The vertically hotter IC stars at
$R\sim R_0$ and $t=t_{\rm f}$ have migrated less than the cooler, old
inserted stars, a result similar to those of both \citet{solway} and
\citet{vera14}. Still, on average IC stars show significant outwards migration over the
$10\gyr$ of the simulation.  (ii) The $\Delta L_z$ distribution for all stars
is less bent than in Y1. This is because of the longer input scalelength of
the oldest populations and the suppressed non-axisymmetric structures at
early times as shown in Figure \ref{barx}.

As shown in Figure \ref{migthin}, Model P1s6 undergoes a very low level of migration to
$R=8\kpc$. This is also clearly visible in Figure \ref{migall}. As it has a
flatter SF history than P2, mass growth is slower at early times and as it forms
inside-out from $1.5$ to $4.3\kpc$ rather than the constant $h_R=2.5\kpc$ in
P2, the density of inserted stars near the centre is lower than in
P2. Thus, non-axisymmetries are strongly suppressed and even the oldest
inserted populations at $R=8\kpc$ and $t=t_{\rm f}$ show hardly any positive
$\Delta L_z$ average and the thick IC population shows essentially no
migration at all. From Figure \ref{migthin}, we have learnt that Model P1s6 shows
too little migration to be a viable representation of the MW. As Model P2 shows the
right amount of radial migration and its thick-disc stars at $R=8\kpc$ and
$t=t_{\rm f}$ are on average outwards migrators, we conclude that this is
likely also the case for thick-disc stars found in the Snhd.

Model \V6 differs in several ways from the previous models, but confirms what we
have concluded so far.  Its IC is a low-mass elliptical and its oldest
inserted stars are hot disc stars. Both show significant outwards migration,
which, due to cooler kinematics, is stronger for the inserted stars. Despite
being born hot, the oldest populations, which make up the thick disc and are
thus the equivalent of the IC stars in the P models, show a $\Delta L_z$
distribution strongly skewed to positive values. Model \V6 has a steeper than average
decline in its SFR ($t_{\rm SFR}=6\gyr$) and an inside-out formation history which
grows from a very compact initial $h_R=1.0\kpc$ to $3.5\kpc$. As shown in
Figure \ref{barx}, it has significant non-axisymmetric structure from the
beginning. Consequently, the $\Delta L_z$ distribution is bent in a similar
way to that of the thin-disc model Y1. As the old, chemically defined thick
disc of the MW is likely more compact than the thick disc in the P models
\citep{bovy,hayden}, the migration behaviour in \V6 is likely a better model for
this class of stars. This means that the chemically defined, old thick-disc
stars in the Snhd have likely been born at significantly lower $L_z$, as
required by chemical evolution models \citep{sb09a}.

The lower row of panels in Figure \ref{migall} deals with the separate issue of 
how the vertical velocity dispersion $\sigma_z$ varies with $\Delta L_z$ and thus, 
if outwards migrators are hotter than stars with $\Delta L_z=0$ and make up the 
thickest populations \citep{sb12, roskar} or if they do not help to thicken the disc
\citep{minchev12, vera14}.

In all models old inserted stars are kinematically hotter than young stars,
either because they have been heated by GMCs or because they were born hot.
We will discuss the strength of heating in Section \ref{sec:agevel}, and here
focus on how $\sigma_z$ depends on $\Delta L_z$ at a given $\tau$. In the
thin-disc model Y1, the value of $\sigma_z$ for young stars is essentially
independent of $\Delta L_z$, whereas in P2 the bar formation event at
$t=7\gyr$ introduces a gradient at $\tau=3\gyr$ causing outwards migrators to
be hotter. In both models,  old outwards migrators are hotter than
stars with $\Delta L_z=0$, both for stars born cold and the stars of the
thick IC disc of P2. The latter show a maximum $\sigma_z$ at $\Delta
L_z\sim1000\kms\kpc$ and are somewhat cooler for higher $\Delta L_z$, for
which also the number drops significantly as indicated by the white density
contours. 

\begin{figure*}
\vspace{-0.3cm}
\hspace{-1.1cm}\includegraphics[width=14cm]{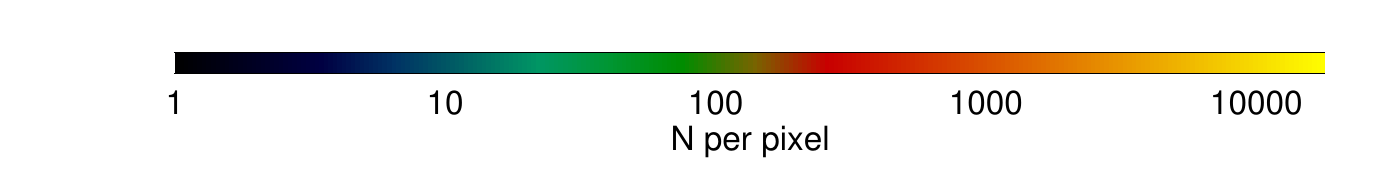}\vspace{-0.3cm}\\
\hspace{-0.1cm}\includegraphics[width=18cm]{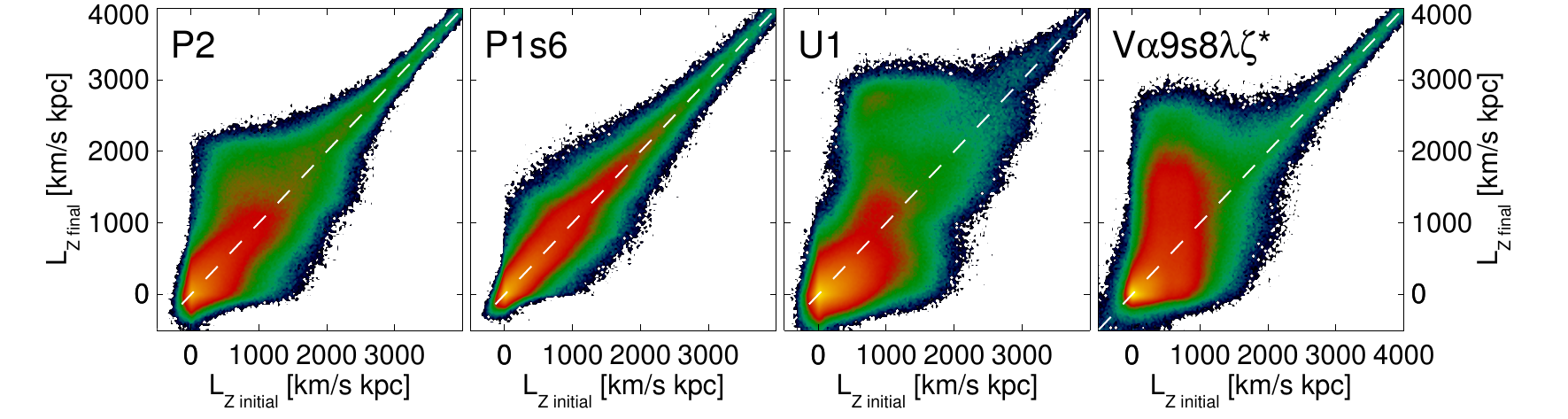}\\
\caption
{The distribution of thick-disc stars in the $L_z(t_{\rm initial})$-$L_z(t_{\rm f})$ plane. For thick-IC disc models
we consider all IC stars at $t_{\rm initial}=0$, whereas for models with declining $\sigma_0$ we consider all stars born
at $t\le2\gyr$ and use $t_{\rm initial}=2\gyr$. The white dashed line marks $L_z(t_{\rm initial})=L_z(t_{\rm f})$.}
\label{migthick}
\end{figure*}

Model P1s6, which has hardly any migration, shows that the non-migrating IC
stars are the hottest IC stars.  Thus for our thick-disc ICs, the strength of
migration determines whether outwards migrators heat the disc, but for
realistic levels of radial migration they do heat it significantly.

The diagram looks quite different for Model \V6. The outwards migrators in the
old, hot population of this model are cooler than the stars with $\Delta
L_z=0$, in contrast to what was found in P2.  For models of this class we
assume a radially constant input dispersion $\sigma_0$ whereas the thick-disc
ICs require an outwards declining vertical dispersion to maintain a constant
scale height.  Stars migrating outwards to lower surface densities and
shallower potential wells will cool as they adiabatically conserve their
vertical actions, whereas stars with $\Delta L_z=0$ will actually undergo
mild adiabatic heating as the surface density increases with time. 

In Model P2 the radial gradient in $\sigma_{z, {\rm thick}}$ caused outwards
migrators to have a higher $\sigma_z$ at the start of the simulation than stars
with $\Delta L_z=0$, and by $t=t_{\rm f}$ adiabatic cooling/heating of the
populations has lowered but could not fully erase this initial difference. In
Model \V6 there was no initial radial gradient in $\sigma_{z, {\rm thick}}$ and
thus outwards migrators have lower $\sigma_z$ than stars with $\Delta L_z=0$.
The fact that, in P2, $\sigma_z$ drops at $\Delta L_z>1000\kpc\kms$ indicates
that the greater tendency of cooler stars to migrate modifies this picture.
We do not know what the initial dispersion gradient in the MW looked like, as
in this work we do not investigate the heating mechanism for the thick disc
and thus we cannot draw conclusions regarding heating of the outer MW disc by
migrators.

\subsection{Migration of thick-disc stars}

The chemically defined thick disc in the MW has recently been studied over an extended volume
($R=3-15\kpc$, $|z|=0-2\kpc$) by \citet{hayden} using stars from the APOGEE survey.
They show that very few stars of the high $\alpha$ sequence are found outside of $R\sim11\kpc$.
Inside this radius, the high-$\afe$ sequence inhabits the same area in the $\afe$-$\feh$-plane.
Chemical evolution models \citep{sb09b} require that these stars must have formed at early times
and at radii significantly inside $R_0$ and thus must have migrated outwards.

In our simulations, we do not model chemical evolution. Moreover, our
galaxies by construction have for the IC disc and for the inserted stars
exponential profiles, which do not have an outer cutoff for SF, as one would
expect to find in real galaxies. Still, we can inspect the extent of radial
migration by thick-disc stars and check (i) whether a significant number of
stars from the inner galaxy have reached the Snhd, and (ii) whether at
$t=t_{\rm f}$ there is a characteristic radius or angular momentum $L_z$,
beyond which one is extremely unlikely to find an outwards migrator from the
early disc.

In Figure \ref{migthick} we therefore plot the distribution of thick-disc stars in the 
$L_z(t_{\rm initial})$-$L_z(t_{\rm f})$ plane. Thick-disc stars and $t_{\rm initial}$ are defined as
follows: (i) for models with thick IC discs we consider all IC stars at $t_{\rm initial}=0$, 
and (ii) for models with declining $\sigma_0$ we consider all stars born at $t\le2\gyr$ and 
use $t_{\rm initial}=2\gyr$. The white dashed line marks $L_z(t_{\rm initial})=L_z(t_{\rm f})$.
Considering a Snhd like region at $R=8\kpc$, the corresponding angular momentum for our models
at $t=t_{\rm f}$ is in the range $L_z(t_{\rm f})=1800-2000\kpc\kms$ for circular orbits.
Note that stars on eccentric orbits spend more time near apocentre than near pericentre
and thus the $L_z$ distribution at a given radius will be skewed towards lower $L_z$.

As discussed in previous sections, the migration in Model P1s6 is suppressed
as the strength of non-axisymmetric structures is low at all times. Thus at
$t=t_{\rm f}$ there are few thick-disc stars at values of $L_z$ typical for
the Snhd. The situation is markedly different in the other three models
shown. All of these models predict a significant number of outwards migrators
in today's Snhd from $L_z(t_{\rm initial})\la500\kpc\kms$. The distribution
of stars with $L_z(t_{\rm f})=1800-2000\kpc\kms$ is skewed towards lower
$L_z(t_{\rm initial})$ and thus outwards migrators in these models.  As
already discussed in Section \ref{sec:migsnhd}, the Snhd is more dominated by
outwards migrators in Model \V6 than in Model P2, because at early times \V6 had both a
more compact thick disc and stronger non-axisymmetries. 

Models P2, U1 and \V6 also show very little migration in the outer disc, as
we find $L_z(t_{\rm initial})\approx L_z(t_{\rm f})$ for $L_z(t_{\rm
initial})\ga3000 \kpc\kms$.  This is because in all models the outer discs
are DM dominated and disc self-gravity is too weak to generate the
non-axisymmetric structures required to drive significant radial migration.

Moreover, these three models show that there is an effective upper boundary
$L_{z, {\rm final, edge}}$ that the stars with low $L_z(t_{\rm initial})$ can
reach. This boundary varies little with $L_z(t_{\rm initial})$ so long as
$L_z(t_{\rm initial})<L_{z, {\rm final, edge}}$. The value of $L_{z, {\rm final, edge}}$ 
does vary from simulation to simulation: we find 
$L_{z, {\rm final, edge}}\approx 2300\kpc\kms$ in P2, $L_{z, {\rm final, edge}}\approx
2800\kpc\kms$ in \V6 and $L_{z, {\rm final, edge}}\approx 3100\kpc\kms$ in U1.
The boundary reflects the absence of significant drivers of radial migration
in the outer disc. $L_{z, {\rm final, edge}}$ is higher in U1 than P2, because
U1 is baryon-dominated to a larger radius than P2 (Figure \ref{vcirc}).  This
upper bound for $L_{z}(t_{\rm final})$ explains why \citet{hayden} find that
the chemically defined thick disc peters out at $R\sim10\kpc$ and confirms
that the outer-most stars of the chemically defined thick disc can be
migrators from the early inner disc despite their being vertically hot.

\section{Solar neighbourhood kinematics}
\label{sec:kinematics}

\begin{figure*}
\vspace{-0.cm}
\includegraphics[width=18cm]{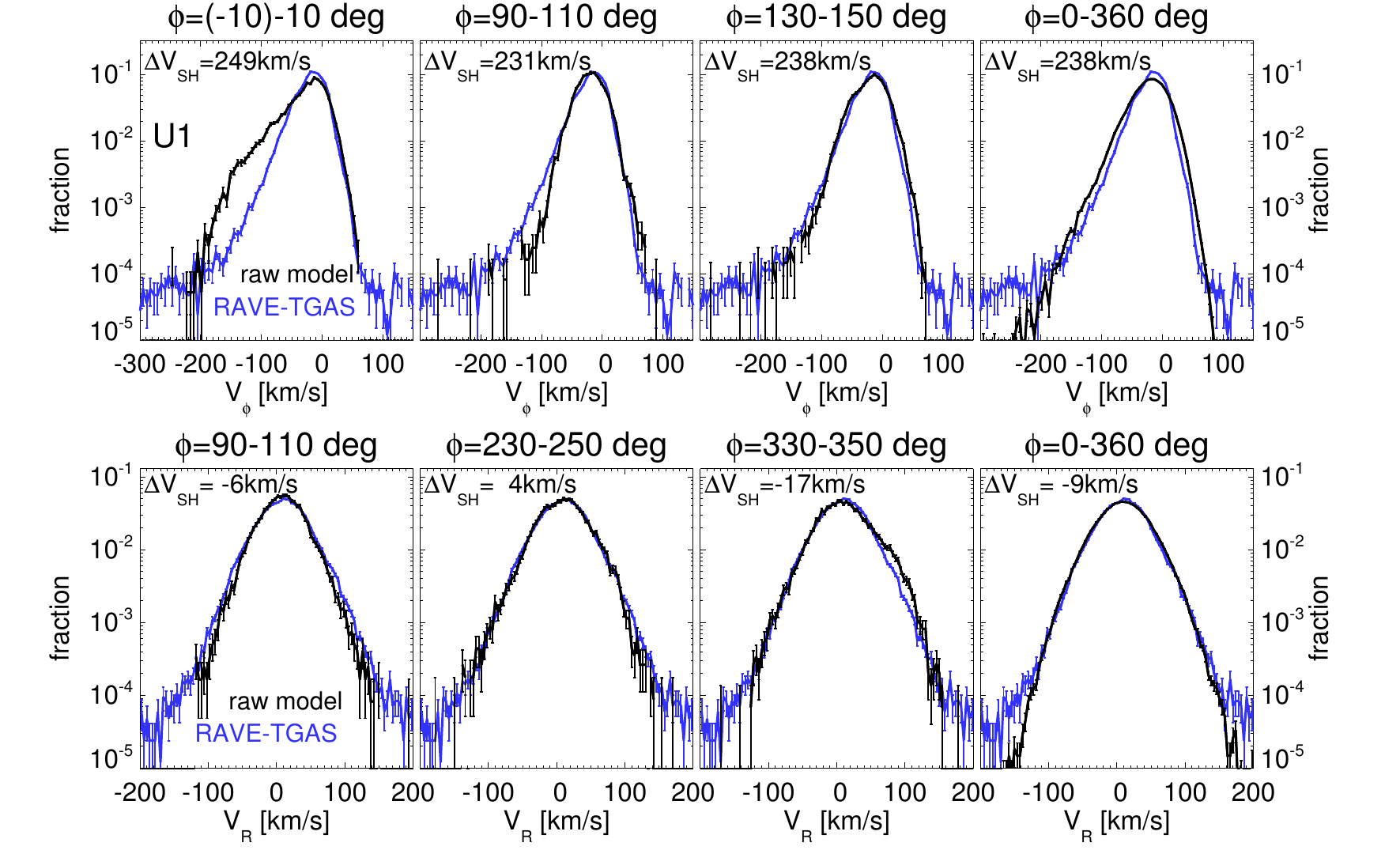}\\
\caption
{Azimuthal variation of in-plane velocity histograms in direction of rotation $V_{\phi}$ (top row)
and radial direction $V_{R}$ (bottom row) in model U1 for three azimuthal bins each (left three columns)
and an azimuthal average (right column). The numbers in the top left corners of each panel
show the relevant component of $\Delta{\bf V}_{\rm sh}$ in $\kms$, the velocity by which the model histograms were shifted. 
Here we consider only altitudes $|z|<250\pc$. Black points are model data and blue 
points are RAVE-TGAS data. Poisson error-bars are shown for all histograms.}
\label{hv-azi}
\end{figure*}

\subsection{RAVE-TGAS velocity distributions}
\label{sec:tgas}

Next we compare kinematics of model stars at different altitudes to data from
the Gaia mission \citep{gaia}. The release of Gaia data includes the
Tycho-Gaia Astrometric Solution (TGAS, \citealp{tgas}), which contains proper
motions and parallaxes $\varpi$ for $\sim 2\times10^6$ nearby stars. Combined
with line-of-sight velocities $v_\parallel$ from the Radial Velocity Survey (RAVE,
\citealp{kunder}), 6D phase space information is available for 
$\sim 2.5\times10^5$ stars. 

\subsubsection{RAVE-TGAS data}

We use the catalogue provided by the 5th data release of the RAVE survey
\citep{kunder}, which contains $255\,922$ observations of stars to which TGAS
counterparts have been assigned.  As the catalogue contains multiple observations
of certain stars, we first look for observations which have identical TGAS proper
motions and parallaxes. For multiple observations of the same object we select 
the one with the smallest observational error in $v_\parallel$. We discard stars with
negative parallaxes and stars with parallax errors $\sigma_{\varpi}>0.25\varpi$.
For simplicity we assume that the distances $s$ of surviving stars can be calculated
as the inverse of the parallaxes $\varpi$, $s=1/\varpi$. At the accuracy required for 
our comparison, this is a reasonable assumption as was shown by
\citet{sa17}.  As the smallest parallax errors in TGAS are
$\sigma_{\varpi, {\rm min}}\sim0.21\;{\rm mas}$, distances are limited to $s\la1.2\kpc$.

For the remaining $\sim 1.3\times10^5$ stars we calculate positions in Galactic 
coordinates and the Cartesian components $U,V,W$ of heliocentric 
velocity, with $U$ directed towards the Galactic centre and $V$ in the direction 
of Galactic rotation.  The  components $U,V$ are then transformed into the 
component $V_R$  from the Galactic centre to the position of the star, and  
$V_{\phi}$ in the direction of a circular orbit at the position of the star. To 
determine $V_R$ and $V_{\phi}$, we require the in-plane velocities of the Sun in 
the Galactic rest frame and distance of the Sun to the Galactic centre, which 
\citet{ralph2012} and \citet{ralph2010} determined as $V_{{\rm g}, \odot}=250\kms$, 
$U_{\odot}=11\kms$ and $R_0=8.3\kpc$. We choose the zero points of $V_R$ and 
$V_{\phi}$ so that at the position of the Sun a star has $V_R=-U$ and $V_{\phi}=V$.
 
We bin the sample's stars into the altitude ranges $|z|<250\pc$, 
$250\pc<|z|<500\pc$ and $500\pc<|z|<800\pc$ and create histograms of all three
velocity components in all three ranges. For simplicity we assume Poisson errors
$\sigma_N=\sqrt{N}$, where $N$ is the number of stars in a velocity bin of
size $\Delta V=6\kms$ for $V_{\phi}$ and $\Delta V=4\kms$ for the other components.

\subsubsection{Azimuthal variation of velocity distributions}

To compare our models to the RAVE-TGAS data, we select stars at
$R=8.3\pm1.0\kpc$ and for them create histograms of $V_z$, $V_R$ and
$V_{\phi}$ at the same three altitude ranges with the same bin sizes $\Delta
V$. To understand the azimuthal variation of the velocity distribution, we
divide the cylindrical shell into 18 equally sized parts, each 20 degrees
wide and additionally create histograms for all azimuths combined. The angle
between the line Sun-Galactic centre line and the major axis of the bar is usually
assumed to be in the range $15-30$ degrees (see e.g. \citealp{binney91,
wegg}) on the trailing side of the bar. We define an angle $\phi$ for our
models, such that the tips of the bar are at $\phi=0$ and $180$ degrees and the
most likely Snhd bins are at $\phi=10-30$ and $190-210$ degrees. 

To illustrate the azimuthal variations of the $V_R$ and $V_{\phi}$
distributions,  we show in Figure \ref{hv-azi} data from model U1, which, as
was discussed in Paper 3, has a strong and unrealistically long bar. Consequently its
distributions of $V_R$ and $V_\phi$ vary significantly with azimuth.  By
contrast its distribution of $V_z$ varies insignificantly, so we do not show
them. 

Given that the Sun has a velocity $U_\odot\simeq11\kms$ towards the Galactic
Centre, the $R$ component of the measured heliocentric velocities of stars
are larger than their Galactocentric velocities by $\Delta V_R\sim11\kms$.
Hence in each panel of Figure \ref{hv-azi} U1's velocity histogram has been
shifted by taking from the Galactocentric velocity a vector $\Delta {\bf
V}_{\rm sh}$. For any given panel, we choose $\Delta {\bf V}_{\rm sh}$ to
optimise the fit between the model histograms and the corresponding
histograms of the velocities of RAVE-TGAS stars, which are plotted in blue.
If the Galaxy were axisymmetric, the $R$ component of ${\bf V}_{\rm sh}$
would be the reflex of the Solar motion with respect to the Local Standard of
Rest (LSR), so $\Delta V_{{\rm sh},R}=-6\kms$ would imply $U_\odot=6\kms$, and
the $\phi$ component of $\Delta{\bf V}_{\rm sh}$ would be the sum $v_{\rm
circ}+V_\odot$ of the local circular speed and the Sun's peculiar azimuthal
velocity. Since model U1 is strongly non-axisymmetric, the shifts required to
optimise the fit to the RAVE-TGAS data vary with azimuth as shown at the top
of each panel of Figure \ref{hv-azi}. These variations reflect the fact that
the non-axisymmetric component of the potential causes the whole Solar
neighbourhood to move in and out and to slow down and then speed up as it
moves around the Galactic Centre.

The values of $\Delta{\bf V}_{\rm sh}$ given in Figure \ref{hv-azi} were
determined as follows.  At a given azimuth $\phi$, we systematically shifted
the model data in steps of $1\kms$ independently in all three velocity
components and used the same binning as applied for the Snhd data. Then
for each component $V_j$ we computed the figure of merit
\begin{equation}
\chi^2= \sum_{i_{\rm min}}^{i_{\rm max}}{\bigl( n_{i,{\rm model}} -
n_{i,{\rm data}} \bigr)^2 \over 
 \sigma_{n,i, {\rm model}}^2 + \sigma_{n,i, {\rm data}}^2},
\end{equation}
where $n_i\equiv N_i/N$ and $\sigma_{n,i}=\sqrt{N_i}/N$ with $N_i$ the number
of stars in velocity bin $i$ and $N$ is the total number of stars at the given
range in azimuths and altitudes. The sum runs between the
minimum and maximum bins considered, the choice of which is not critical as
long as the majority of stars are included because the bins in the wings of the
distributions have low $N_i$ and thus contribute little to $\chi^2$.
The values of $\Delta V_{\rm sh}$ given at the top of each panel of Figure
\ref{hv-azi} are those that minimise $\chi^2$.  

The first row in Figure \ref{hv-azi} shows $V_{\phi}$ distributions in model
U1 at $|z|<250\pc$ for three different azimuthal bins and averaged over all
$\phi$. At an angle $\phi$ aligned with the major bar axis in the range
$(-10,10)$ degrees, the $V_{\phi}$ distribution of the model is much more skewed towards
low $V_{\phi}$ than is that of the Snhd data, while in the range $(90,110)$ degrees,
the model distribution drops more sharply at low $V_{\phi}$. In the range
$(130,150)$ degrees, the model agrees reasonably with the Snhd data, but this
azimuthal range is on the leading side of the bar.  We note that the steep decline
towards high $V_{\phi}$ is similar at all azimuths and agrees with the data,
which is why our algorithm chooses $\Delta V_{{\rm sh},\phi}$ so the histograms
agree roughly in this $V_{\phi}$ range.  When all azimuthal bins are considered,
$\Delta V_{{\rm sh},\phi}$ varies between $226$ and $249\kms$. The top right panel
shows that when we average over all azimuths, the model $V_{\phi}$
distribution is similar in shape but wider than the data distribution and we
find an intermediate $\Delta V_{{\rm sh},\phi}=238\kms$. The width of the
distribution is determined by two factors: (i) the varying widths at different
azimuths and (ii) the variation of $\Delta V_{{\rm sh},\phi}$, which broadens the
distribution.

The second row in Figure \ref{hv-azi} shows the $V_{R}$ distributions of model
U1. The variation with $\phi$ is less pronounced than in the case of $V_{\phi}$. At
$\phi=90-110$ degrees the distribution is narrowest, as for $V_{\phi}$. Close to
the major bar axis (here $\phi=330-350$ degrees), we find a distribution
which is skewed towards high $V_{R}$, whereas at $\phi=230-250$ degrees there
is reasonable agreement with the Snhd data. Considering all azimuthal bins,
$\Delta V_{{\rm sh},R}$ varies between $-23$ and $+4\kms$. The azimuthally
averaged histogram is again widened around $V_R\sim0$ because of this
systematic variation, but agrees reasonably well with the data.

We note that none of the model histograms show enough stars at the extreme
wings of the Snhd velocity distributions.  This is not worrying, as (a)
observational velocity errors were not modelled and (b) halo stars are not
included in the model. See \citet{ralph2011} for an illustration of how both
points alter the $V_{\phi}$ distribution.

Figure \ref{hv-azi} clearly shows that it is essential to choose an
appropriate azimuth $\phi$ when comparing a model with non-axisymmetries to
Snhd data. Both using an azimuthal average and choosing a wrong azimuth can
lead to significantly misleading conclusions.

For most models, we find that $\Delta{\bf V}_{\rm sh}$ is not constant with
altitude.  At a Snhd-like value of $\phi$, $\Delta V_{{\rm sh},\phi}$
for most models decreases with increasing $|z|$ by up to $\sim10\kms$; only in a
very few models does $\Delta V_{{\rm sh},\phi}$ increase with $|z|$.  Most
models show altitude variations of $\Delta V_{{\rm sh},R}$ smaller than
$\pm3\kms$. These variations are not significant as the Poisson errors are large 
at higher latitudes.  The strongest deviations are found in models with strong bars,
such as U1, where we find $\Delta V_{{\rm sh},R}$ to be lower by $\sim8\kms$
at higher $|z|$.  These variations with altitude are in part
caused by differences in the shape of the velocity distribution between model
and data.  Uncertainties in the observations, which increase with distance
and vary with Galactic latitude are also expected to play a role.

$\Delta V_{{\rm sh},z}$ usually shows very little
variation with either azimuth or altitude.  Typical values are in the range
$\Delta V_{{\rm sh},z}=7-8\kms$ in agreement with LSR determinations for the Snhd
(e.g. \citealp{db98, ralph2010}). In model U1 we find a mild systematic
variation with altitude with lower shifts near the bar tips, $\Delta V_{{\rm sh},z}=6-7\kms$, 
and higher shifts at $\phi\approx90-130$ and $270-310$ degrees,
$\Delta V_{{\rm sh},z}=8-9\kms$.

\begin{figure*}
\vspace{-0.cm}
\includegraphics[width=16cm]{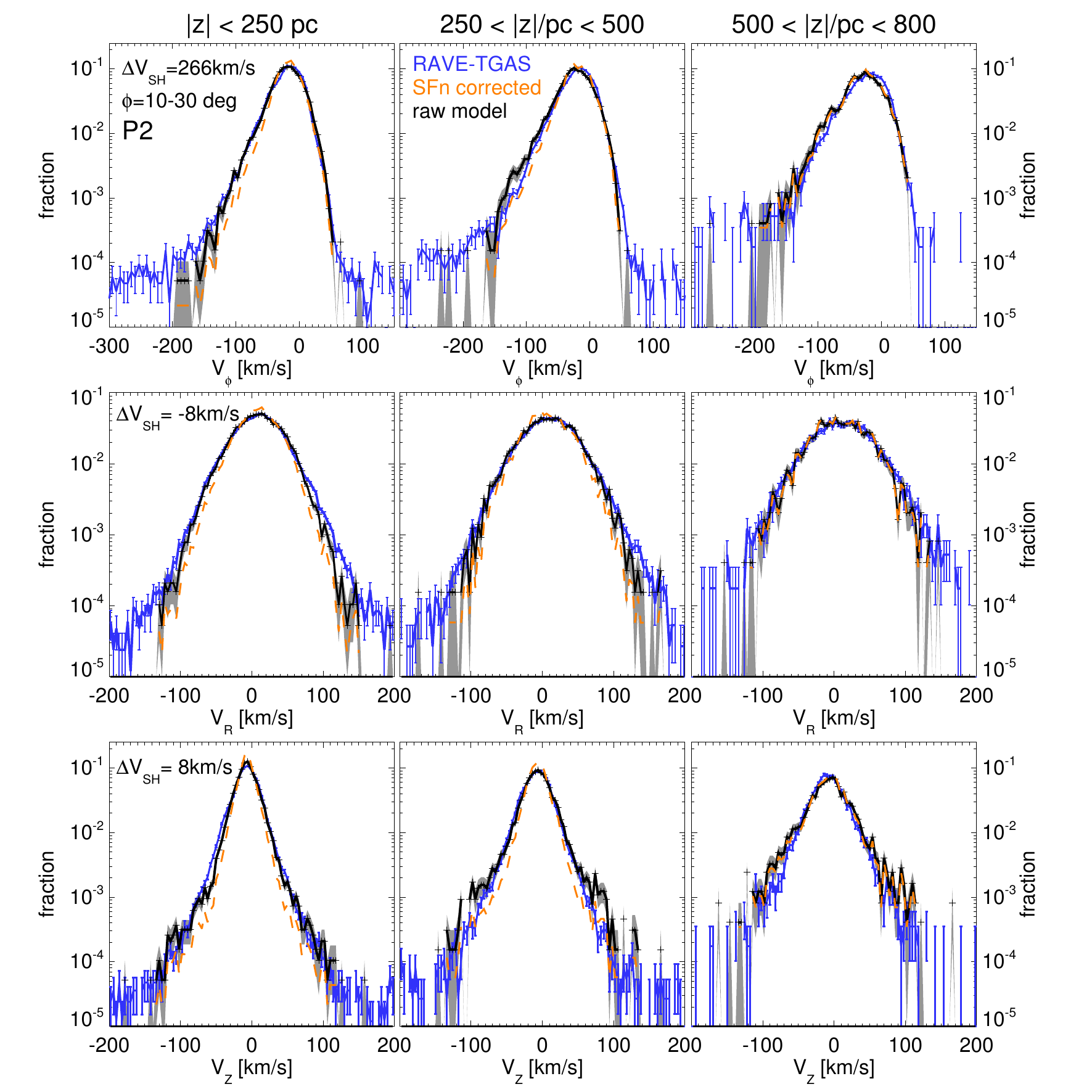}\\
\caption {A comparison of velocity distributions in model P2 with RAVE-TGAS
data. Each row shows a different component: $V_{\phi}$ top row, $V_{R}$
middle row and $V_{z}$ bottom row.  Each column shows a different altitude
bin: $|z|<250\pc$ (left), $250<|z|/\pc<500$ (middle) and $500<|z|/\pc<800$
(right). Snhd data with Poisson error-bars are shown in blue. Raw model
histograms are shown in black with grey shaded areas marking the Poisson
uncertainties. Model histograms altered by the selection function are shown
as orange dashed lines. The number in the top left corner of each leftmost
panel is the relevant component of the shift vector $\Delta{\bf V}_{\rm sh}$.
The latter is determined for raw histograms.  The azimuthal range
considered is at an angle $\phi=10-30$ degrees to the major axis of the bar
on the trailing side, similar to the Snhd. The radial range considered for
the models is $R=8.3\pm1.0\kpc$.} \label{hv_P2}
\end{figure*}

\subsubsection{Selection function}

When comparing our models to Snhd data we should consider the effects of
survey selection functions (SFns), as was discussed in Paper 2 for the
age SFn of the GCS data (see also Section \ref{sec:agevel}). Here we only seek 
a qualitative understanding of the age SFn of our sample. To circumvent the 
spatial SFn (see \citealp{sa17}), we consider stars at different altitudes 
$|z|$ as explained above. We do not expect significant variations of stellar 
kinematics over distances $s\la1\kpc$ in the $R$ and $\phi$ directions.

For our spatially limited RAVE-TGAS sample as selected above, we consider the
$I$-band magnitude selection of RAVE stars in the interval $9<I<12$
\citep{wojno} to  dominate. To model the SFn, we employ the
population synthesis machinery from the \citet{sb09a} model 
as described in Section 2 of \citet{as15}. We find that for all distances
$s\la1\kpc$, the selection probability $p$ is well approximated by
$p\propto {{1}/(\tau / \gyr +1.5)}$, where $\tau$ is the age of the star.
 
To calculate SFn weighted velocity distributions, we replace the numbers of
model particles in velocity bin $i$, $N_i$, with the effective number $N_{i, {\rm
eff}}=\sum_{j}{w_j}$, where $w_j={{1}/(\tau_{j}/ \gyr+1.5)}$ is the SFn weight
of star $j$. Poisson errors are calculated as 
\begin{equation}
\sigma_{N_{i, {\rm eff}}}={{\sum_{j}{w_j}} \over { \sqrt{\sum_{j}{w_j^2}}}}.
\end{equation}

The effect of this selection can be understood from Figure \ref{hv_P2}. The
right column shows the velocity distributions of stars in model P2 at the
highest altitude bin $500<|z|/\pc<800$, which is dominated by the thick-disc
stars, which in this model come from the IC. Since the vast majority of those
stars are old, there is little difference between the intrinsic distributions
(black) and the SFn weighted distributions (orange). At lower altitudes,
shown in the left and middle columns, the impact of the SFn is clearly visible. The
SFn produces narrower velocity distributions for all velocity components
because it prefers younger and thus kinematically cooler stars. The effect is
somewhat stronger at intermediate altitudes, as they contain a more equal
mixture of younger and older stars, whereas the lowest altitude bin has a
high fraction of young stars. The distributions in $V_{\phi}$ are asymmetric
and the SFn hardly affects the steep falloff towards higher velocities. As
was e.g. shown in figure 2 of \citet{sb12}, hotter populations have their
distributions strongly skewed towards lower $V_{\phi}$ and bins at high
$V_{\phi}$ are thus always dominated by young and cool populations, so the
SFn has little effect at high $V_{\phi}$.

\subsubsection{Model P2}

In Figure \ref{hv_P2} we compare RAVE-TGAS data with the velocity distributions
of Model P2 in one of the two possible Snhd locations:
$R=8.3\pm1.0\kpc$ and $\phi=10-30$ degrees. As the bar has $m=2$
symmetry, we can find locations near either end of the bar and choose the one
that agrees best with the Snhd data. One has to keep in mind that we are not
in any way fitting to the data.  Model P2 existed before the TGAS data were
published and was created with structural and AVR constraints in mind.  The
RAVE-TGAS data are thus an independent evaluation of our models. As was
already noted, we do not model the effect of halo stars and observational
errors and thus do not expect agreement between model and data in the extreme
wings of the distributions. For model P2, the discrepancies appear roughly at
$|V_R|>100\kms$, $|V_z|>120\kms$, $V_{\phi}>50\kms$ and $V_{\phi}<-150\kms$.

The data for model P2 agree best with the observations when the SFn is not
taken into account (black points and grey shaded error regions). Indeed then
for all velocity components the model agrees quite well with the observations
at all altitudes. Applying the SFn (orange dashed line) generally makes
the model velocity distributions too narrow except  in the highest altitude bin,
where the SFn has no significant impact. 

\begin{figure}
\vspace{-0.cm}
\includegraphics[width=8cm]{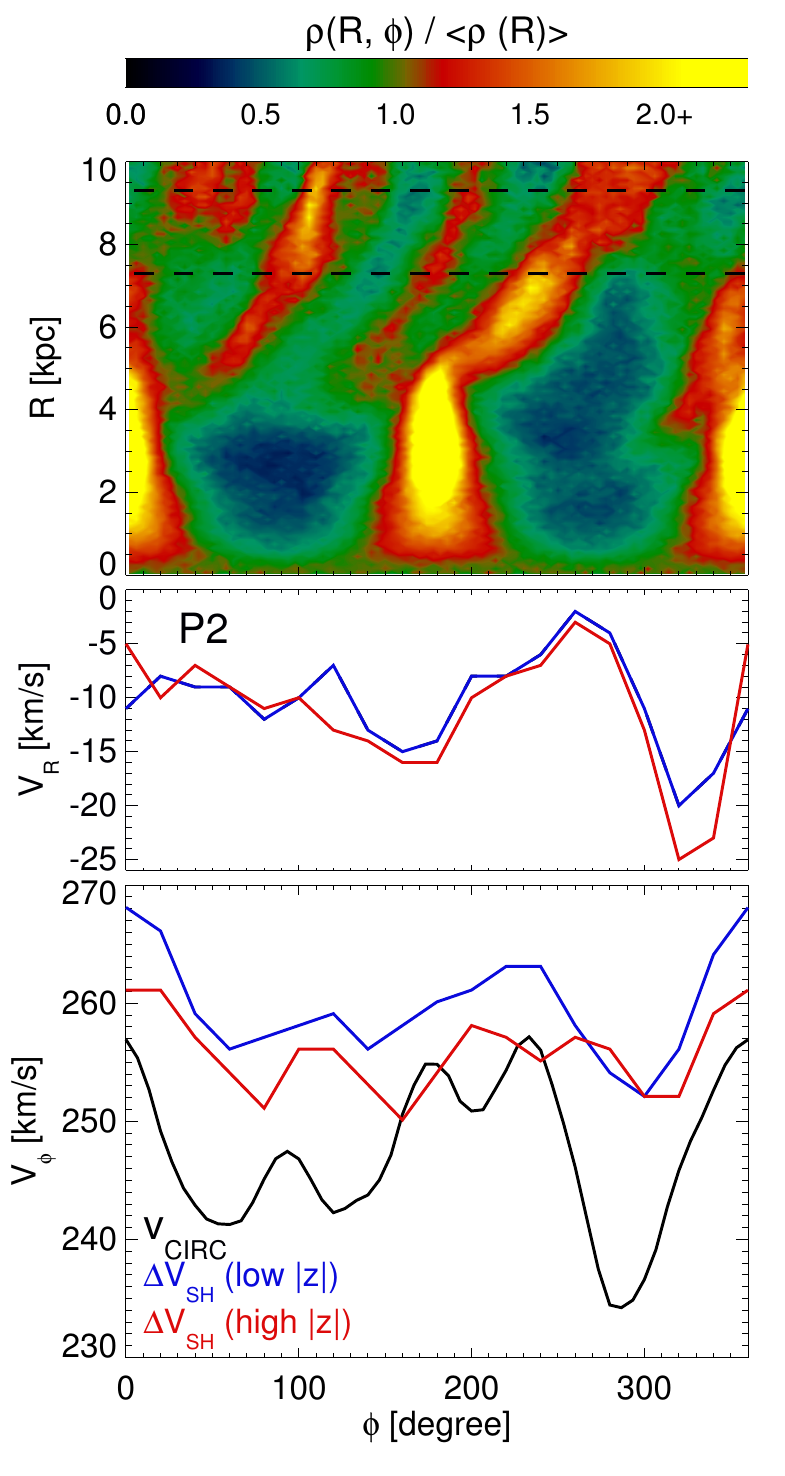}\\
\vspace{-0.5cm}
\caption
{Bottom panel: the azimuthal variation at $R=8.3\kpc$ in model P2 of the circular speed $v_{\rm circ}=\sqrt{{a_R(R)} R}$ (black), 
the $\phi$ component of the velocity shift $\Delta{\bf V}_{\rm sh}$ found for altitudes $0<|z|/\pc<250$ (blue) and for $500<|z|/\pc<800$ 
(red). Middle panel: azimuthal variation of the $R$ component of the velocity shift. Top
panel: non-axisymmetries in P2 in the $R$-$\phi$-plane 
as traced by the overdensity $\rho(\phi,R) / \langle \rho(R) \rangle$. Dashed lines mark the region
$R=8.3\pm1.0\kpc$, used for velocity distributions.}
\label{vcv}
\end{figure}

Considering the raw $V_{\phi}$ distributions in the relevant regions, we find
good agreement at all scaleheights.  At $500<|z|/\pc<800$, the centre of the
distribution is shifted to lower $V_{\phi}$ than in the data. As already
noted, if our algorithm is allowed to choose separate $\Delta{\bf V}_{\rm sh}$ 
at different altitudes, for most models it chooses lower $\Delta V_{{\rm sh},\phi}$
 at higher $|z|$, in this case $5\kms$ lower. At $250<|z|/\pc<500$,
a $3\kms$ difference in $\Delta V_{{\rm sh},\phi}$ would allow for an even
better fit.  At low $|z|$, and for all three altitude bins combined, we find
$\Delta V_{{\rm sh},\phi}=266\kms$. We should compare this to the Galactic
rest frame velocity component in rotational direction $V_{{\rm g},
\odot}=250\pm9\kms$ as found by \citet{ralph2012}. This paper suggests that
$V_{{\rm g}, \odot}$ comprises a local circular speed $v_{{\rm
circ}}=238\pm9\kms$ and a motion relative to the LSR $V_{\odot}=12\pm2\kms$.
From Figure \ref{vcirc}, we learn that the azimuthally averaged $v_{{\rm
circ}}\approx247\kms$ is at the upper end of the range allowed for the Snhd,
which leaves a discrepancy of $\sim7\kms$ for $V_{\odot}$. As already noted,
$\Delta V_{{\rm sh},\phi}$ varies with azimuth $\phi$, for this model at
$|z|<250\pc$ between $252$ and $268\kms$ as shown by the blue line in the
lower panel of Figure \ref{vcv}.  The other possible Snhd location
($\phi=190-210$ degrees) has $\Delta V_{{\rm sh},\phi}=261\kms$ and a
narrower $V_{\phi}$ distribution. 

Figure \ref{vcv} shows that, along the direction of one bar tip, $\Delta V_{{\rm sh},\phi}$ 
has a maximum and the minima lie close to minor axis. Another maximum lies not along 
the other bar tip, as one would expect if the bar was completely dominant, but at
$\phi\approx240$ degrees. To understand this better, the black curve in the
lower panel of Figure \ref{vcv} shows the azimuthal variation of $v_{\rm
circ}\equiv\sqrt{{a_R(R)} R}$.  We calculate $v_{\rm circ}$ at three radii
$R=7.8$, $8.3$ and $8.8\kpc$ and $54$ equally spaced azimuths $\phi$ each.
Then we average over 3x3 points each to reduce $N$-body noise and determine
$v_{\rm circ}(\phi, R=8.3\kpc)$, which we find varies between $234$ and
$256\kms$ and thus by slightly more than $\Delta V_{{\rm sh},\phi}$. The
curves are offset by $\sim10-15\kms$ because $V_{\odot}=12\pm2\kms$, but the
positions of the extrema agree well. The variations are smaller in $\Delta V_{{\rm sh},\phi}$
than $v_{\rm circ}$ because the stars have a non-negligible
velocity dispersion. The red line shows $\Delta V_{{\rm sh},\phi}$ for
$500<|z|/\pc<800$, a region dominated by kinematically hotter thick-disc
stars. Consequently, the variations are smaller than at $|z|<250 \pc$, where
thin-disc stars dominate.  As noted above, at higher $|z|$ the model stars
lag the thin-disc stars more than in the Snhd, so the red curve lies at lower
$V_{\phi}$ than the blue curve. 

To understand the deviation from a simple $m=2$, bar-dominated picture, we
show in the upper panel of Figure \ref{vcv} the azimuthal stellar density
variation due to non-axisymmetries, tracked by the fractional azimuthal
variation $\rho(\phi,R) / \langle\rho(R)\rangle$ in the $R$-$\phi$ plane.
Clearly, at $R<5\kpc$, the bar dominates, as we can also learn from
Figure \ref{sd}.  Outside $R=5\kpc$ and in the region $R=8.3\pm1.0\kpc$, from
which we select stars for our velocity histograms, a four-armed spiral
pattern is visible, which itself shows significant substructure. The density
peaks and troughs clearly correlate with the structure in the $v_{\rm
circ}(\phi)$ curve -- for example the bar-related minimum at
$\phi\approx280$ degrees is enhanced by a spiral arm which at these azimuths lies
beyond $R=8\kpc$. We have no reason to believe that the model's spiral
structure provides a close match to the Galaxy's spiral structure -- we have
shown that P2's bar has a reasonable length but have not shown that its
spiral structure resembles that of the Galaxy. Hence we should not expect
the model curves to reproduce the observations in more than general
characteristics.

Figure \ref{vcv} suggests that non-axisymmetric structures cause the
considered region in model P2 to move at $\Delta V_{\phi}\sim7\kms$ relative
to the average circular velocity. We note that \citet{bovy12} find 
$v_{{\rm circ}}=218\pm6\kms$ and $V_{{\rm g}, \odot}=242^{+10}_{-3}\kms$ for the Snhd
and attribute the large difference $\sim25\kms$ to the sum of
$V_{\odot}\sim 12\kms$ and a systematic motion of the Snhd relative to the
average circular velocity at $R_0$ of order $\Delta V_{\phi}\sim 10-15\kms$. 
This proposed difference is even larger than that found in model P2.

In the altitude bin $|z|<250\pc$, which is dominated by the thin disc, the
radial and vertical velocity distributions are both somewhat too narrow, even
when the SFn is neglected.  This is interesting because P2's thin-disc
scaleheight, $h_{z, {\rm thin}}\approx210\pc$, suggests an unrealistically
narrow $V_{z}$ distribution (see Figure \ref{verts}), but Paper 1 suggests
that the appropriate level of migration and bar length should correspond to
an appropriate radial velocity dispersion.

At higher $|z|$, the distributions of $V_z$ when the SFn is neglected are
slightly too broad, whereas the corresponding $V_R$ distributions are slightly
too narrow. P2 at higher altitude is dominated by the thick IC stars, which
were set up with $\sigma_R/\sigma_z=1$. The comparison with the Snhd data
suggests that $\sigma_R>\sigma_z$  might be more appropriate.

An offset $\Delta V_{{\rm sh},z}=8\kms$ is in agreement with the solar
peculiar velocity $W_{\odot}=7.3\pm1\kms$ found by \citet{ralph2010}. We
similarly find $\Delta V_{{\rm sh},R}=-8\kms$. Bearing in mind that $V_R=-U$,
this is only slightly inconsistent with the solar peculiar motion
$U_{\odot}=11\pm1\kms$ found by \citet{ralph2010}. For the RAVE-TGAS sample
of \citet{sa17}, $U_{\odot}\sim10\kms$ is favoured. We note that we are not
expecting to be able to determine $U_{\odot}$ precisely with our method, as
the detailed shapes of the $V_R$ velocity distributions differ between model
and data. Additionally, there is a likely connection between $-\Delta V_{{\rm sh},R}$
being lower than $U_{\odot}$ and the streaming motions of this region in
the model relative to a hypothetical circular orbit, as discussed above for
$V_{\phi}$ and $V_{\odot}$. Indeed, the middle panel of Figure \ref{vcv}
shows that at both low and high $|z|$, $\Delta V_{{\rm sh},R}$ varies with
$\phi$ between $-20$ and $-2\kms$.

\begin{figure*}
\vspace{-0.cm}
\includegraphics[width=16cm]{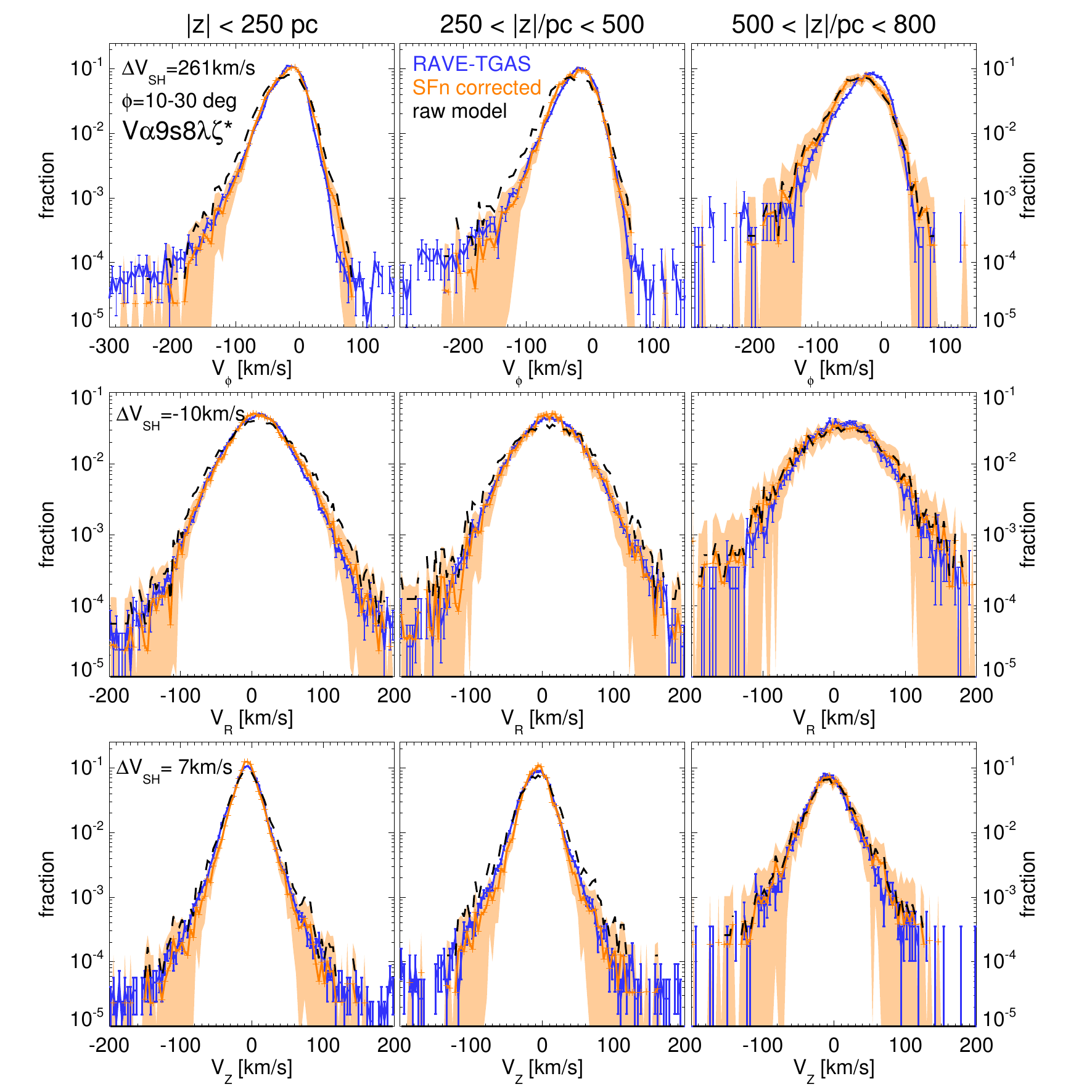}\\
\caption {Same as Figure \ref{hv_P2} but for model \V6. Distributions
considering SFns are shown as orange points with shaded areas showing the
Poisson uncertainties. Dashed black lines are for not taking the SFn into
account. The shift vector $\Delta{\bf V}_{\rm sh}$ was determined using
histograms which take into account the SFn. The components of $\Delta{\bf V}_{\rm sh}$
are shown in the top left corners of the first row panels.} \label{hv_V6}
\end{figure*}

\subsubsection{Model \V6}

In Figure \ref{hv_V6}, we compare velocity distributions from Model \V6 with
the RAVE-TGAS data. We chose \V6 as a counterpart to P2 because it is a
model for which the intrinsic velocity distributions (black dashed lines)
are generally too wide at low and intermediate $|z|$, whereas the SFn
adjusted distributions (orange points with shaded areas for Poisson errors)
show good agreement overall.  This difference arises because: (i) the thin
disc in \V6 is thicker than that in P2, (ii) its vertical velocity
dispersions are generally higher than those in P2 (Figure \ref{migall}) and
(iii) its thick disc is radially hotter because disc stars were inserted with
$\lambda=\sigma_R/\sigma_z=1.25$, whereas P2's IC thick disc was set up with
$\lambda=1$.

As in the case of  P2, out of the two possibilities, we have chosen the Snhd-like
location which shows the better agreement with data. We note that the stars
of the low-mass elliptical IC used for \V6 are included in the histograms.
At $R=8.3\pm1.0\kpc$ these form a very low density, halo-like component and
add to the wings of the velocity distributions, but they are not an
appropriate halo model for the Snhd as can be seen at low $V_{\phi}$.

If we consider the SFn adjusted $V_{\phi}$ distributions of \V6, we find a
mild overproduction of stars with $V_{\phi}\approx60\kms$ at low $|z|$, good
agreement at intermediate $|z|$ and mild tension at high $|z|$. This tension
results both from a somewhat broader $V_{\phi}$ distribution in the model and
a lower $V_{\phi}$ for the peak of the distribution. The latter is reflected
in the fact, that, for $500<|z|/\pc<800$ only, we find a value of 
$\Delta V_{{\rm sh},\phi}$ that is lower by $9\kms$, suggesting that the thick disc in
the model lags the thin disc more than is the case in the Snhd.

At the chosen location we find $\Delta V_{{\rm sh},\phi}=261\kms$. At the
other possible Snhd location we find $\Delta V_{{\rm sh},\phi}=260\kms$ and a
slightly broader $V_{\phi}$ distribution. $\Delta V_{{\rm sh},\phi}$ varies between
$253$ and $262\kms$ for all azimuthal bins, a smaller amplitude of variation
than in P2. The azimuthal variation in $v_{\rm circ}$ is between $243$ and
$252\kms$, so that our conclusions regarding the azimuthal variation of
$V_{\phi}$ average velocities and the LSR are the same as for P2.

Taking into account the SFn, The $V_R$ distributions of model \V6 agree very
well with the RAVE-TGAS data at all values of $|z|$ considered here. The $V_z$
distributions of \V6 are slightly narrower at lower and intermediate $|z|$.
For the highest $|z|$ bin, we find good agreement for both $V_z$ and $V_R$.
These altitudes are dominated by thick-disc stars, which in this model were
fed to the model galaxy with hot birth dispersions and the assumption
$\lambda=\sigma_R/\sigma_z=1.25$. This choice of the ratio appears to be more
appropriate than $\lambda=1$, which was used for the thick-disc IC of model
P2. Combining the information from models P2 and \V6 and the RAVE-TGAS data,
we can still confirm that $\lambda$ for the thick disc is significantly lower
than the $\lambda\approx1.7-2.5$ typically found for the thin disc (see Paper
2), as was discussed by \citet{piffl}, who argued for $\lambda\approx1$ in
the MW thick disc.

$\Delta V_{{\rm sh},z}$ is stable at $7\kms$ for model \V6, and for
$V_R$ we find $\Delta V_{{\rm sh},R}=-10\kms$ in the Snhd like location shown in
Figure \ref{hv_V6}. Both numbers are in in agreement with the vertical LSR
velocities $W_{\odot}=7\pm1\kms$ and $U_{\odot}=11\pm1\kms$ found by
\citet{ralph2010}.The variation of $\Delta V_{{\rm sh},R}$ with $\phi$
in \V6 is between $-17$ and $-4\kms$, again a smaller variation compared to
P2 and generally underlying the conclusions drawn from model P2.

\subsection{Age Velocity Dispersion Relations}
\label{sec:agevel}

We have now established that the overall velocity distributions in our models
with realistic bars and appropriate levels of migration closely reproduce
Snhd velocity distributions. However, it is hard to quantify which models are
best on account of uncertain selection effects. 

The AVRs extracted from the GCS by \cite{nordstrom} provide an additional
constraint on Snhd kinematics. The GCS is restricted to stars at small
distances $s\la100\pc$, a volume that is under-represented in RAVE-TGAS.
Paper 1 showed that the observed radial and vertical AVRs are reproduced
by models that lack a thick disc but have a thin disc and a dark halo with
appropriate masses and the right quantity of GMCs. Paper 2 showed
further that the SFn of the GCS can hide an old thick-disc population
because the GCS contains predominantly young stars and its age errors are
significant.  Whereas the models discussed in Papers 1 and 2 did not have
thick discs, the models discussed here do. 

\subsubsection{Extracting AVRs from data and models}

\begin{figure*}
\vspace{-0.5cm}
\includegraphics[width=15cm]{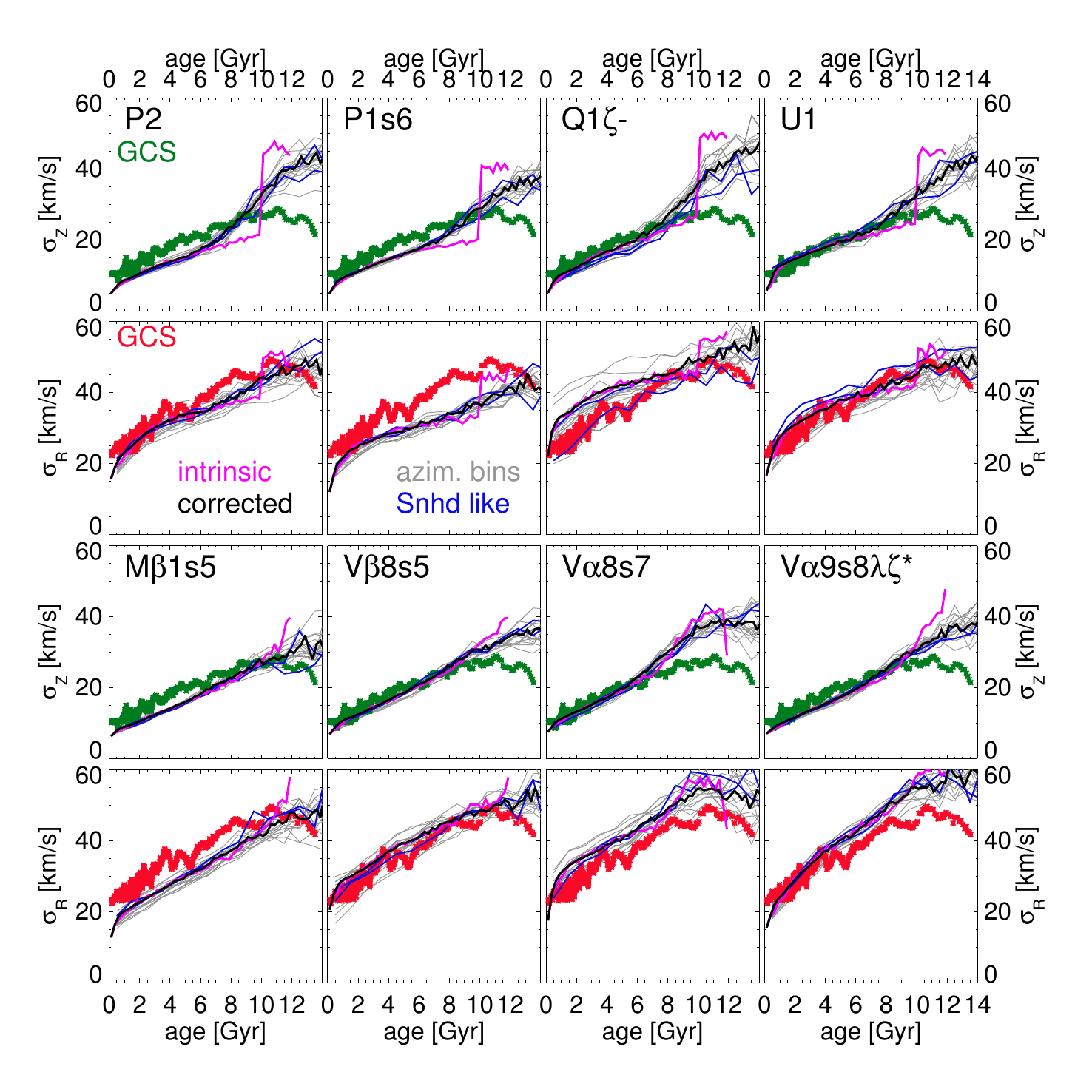}\\
\vspace{-0.5cm}
\caption
{Radial and vertical AVRs in various models. The first and third rows show $\sigma_z(\tau)$
and the second and fourth rows show $\sigma_R(\tau)$. Green and red points are data extracted from \citet{casagrande11}
as described in the text. Black curves are model AVRs at $R=8.3\pm0.5\kpc$, $|z|<100\pc$ and $t=t_{\rm f}$ adjusted for 
age bias and errors. Grey curves are the same for 18 equally spaced azimuthal bins of the same spatial region. Blue
lines are for azimuthal bins which lie in Snhd-like positions relative to the bar. The pink lines show the intrinsic
AVRs without correction for age bias and errors for all azimuths.}
\label{agevel}
\end{figure*}

The green ($\sigma_z$) and red ($\sigma_R$) points in Figure \ref{agevel} are
the AVRs yielded by the ages and velocities of GCS stars given in
\citet{casagrande11}. As detailed in Paper 2, these AVRs are obtained by
first excluding stars with halo characteristics and then selecting stars with
`good' age determinations. The remaining $\sim7\,500$ stars are sorted by age
and the velocity dispersions $\sigma_i(\tau)$ of groups of 200 adjacent stars
are computed, with a new group being formed after moving ten stars down the
rank. Hence every 20th value of $\sigma_i(\tau)$ is statistically independent of its
predecessors.

To determine the AVRs of a model, we select stars satisfying $R=8.3\pm0.5\kpc$
and $|z|<100\pc$. As detailed in Paper 2, these stars are then sorted in age
$\tau$ and assigned weights $w(\tau)$, so that their weighted age
distribution agrees with that of the GCS sample. For models with thick-disc
ICs, which all have $t_{\rm f}=10\gyr$, stars in the ICs are assigned ages
$\tau_{\rm IC} \in (10,12)\gyr$. A model in which $\sigma_0$ declines is
considered only if $t_{\rm f}=12\gyr$, and in such a model we exclude stars
from the low-density elliptical ICs as being `halo' stars. Finally, we
simulate the impact of age errors by scattering the age of each  star
through a Gaussian distribution with dispersion $\sigma_{\tau}=0.2\tau$.

The black lines in Figure \ref{agevel} are for all stars in the selected
volume. The grey lines are for subsets made by dividing the volume in 18
azimuthal bins of equal width. The blue lines are for the bins at
$\phi=10-30$ and $190-210$ degrees and thus at locations relative to the bar
similar to that of the Snhd.  The pink lines show the intrinsic AVRs, before correction
for age bias and errors, and averaged over all azimuths.

\subsubsection{AVRs in thin+thick disc models}

We start by discussing the P models, which live in dark haloes with
concentration parameter $c=9$. Paper 1 showed that in such haloes
thin-disc models (Y models) with a mass of $M_{\rm f}=5\times10^{10}\msun$
reproduce the GCS data quite well.  The pink curves in Figure \ref{agevel}
show that at $\tau=10\gyr$, the thick-IC disc of a P model generates steps in
the pink curves for both the intrinsic $\sigma_z$ and the intrinsic
$\sigma_R$, with the step in $\sigma_z$ being more pronounced. After
correction for age bias and observational errors, the radial AVR of Model P2
(black curve) is still a reasonable match to Snhd AVR, but only on account of
the presence of thick-IC stars. The corrected vertical AVR does not fit the
data well: in the model $\sigma_z$ is too large at $\tau>10\gyr$ and too
small at $\tau<7\gyr$.  The vertical AVR of Model P1s6 is similar to that of
Model P2, but the radial AVR is too low at all ages. 

So both these P models point to inefficient radial heating, especially at
early times.  In Model P1s6, the shortfall in $\sigma_R$ is caused by a lack
of non-axisymmetric structure that was already discussed in Section
\ref{sec:mig}. Model P2, which at $t=t_{\rm f}$ has a more compact stellar
disc than P1s6, shows a bar and a four-arm spiral pattern (see Figure
\ref{vcv}) and sufficient radial migration over the last $5\gyr$. The
non-axisymmetries, however, emerge relatively late (see Figure \ref{barx}),
explaining the low values of $\sigma_R$ for old thin-disc stars. As regards
vertical AVRs, the low dispersions of young stars are associated with relatively
low thin scaleheights $h_{z,\rm thin}\approx 205\pc$ in both P2 and P1s6 (see
Figure \ref{verts}).  In fact, the intrinsic vertical dispersions of the thin-disc
stars are cooler than those in corresponding thin-disc only models at all
ages.

Several factors contribute to the unrealistically low values of $\sigma_z$ in
P models:
\begin{itemize} 
\item[(i)] The Y models presented in Paper 2 do not all reproduce the Snhd
AVRs well -- Y2, the model corresponding to P2, does show slightly lower
$\sigma_z(\tau)$ at almost all $\tau$ as shown in Figure 2 of Paper 2.
The IC disc masses of P models are three times higher than those of Y models,
and at a given final mass $M_{\rm f}$ the SFRs are thus lower. Consequently,
for a given value of $\zeta$, a P model has fewer GMCs, and thus less vertical
heating. 

\item[(ii)] Paper 2 showed that the efficiency of GMC heating depends
on the mass fraction of GMCs. Due to the additional mass in the ICs, the GMC
mass fraction is lower in P models and heating is reduced. 

\item[(iii)] In Figure \ref{migall},
we showed that when the vertical dispersions are determined by GMC
heating, stars that have migrated outwards show higher dispersions than
non-migrated stars. As migration levels in P1s6 are suppressed and in P2 are
only high at late times, $\sigma_z$  is lower than in
corresponding Y models, especially for old stars. 

\item[(iv)] Paper 1 showed that clustering of GMCs in
spiral structures has a mild, but strengthening effect for vertical disc
heating. As structure is suppressed in P models, this also weakens vertical
heating.
\end{itemize}

Consider now Model M$\beta$1s5, which has declining $\sigma_0$, $M_{\rm
f}=5\times10^{10}\msun$ and lives in a dark halo with a high concentration
parameter $c=9$. In this model $\sigma_R$ and $\sigma_z$ are both low at
young ages notwithstanding the model's larger thin-disc scaleheight,
$h_{z,\rm thin}\approx 270\pc$. Its non-axisymmetric structure is too weak,
so it shows too little radial migration.  These characteristics arise from
too little disc self gravity.  Since the circular speed curves of Figure
\ref{vcirc} of models with $c=9$ already have $v_{\rm circ}(R_0)$ at or above
the upper allowed limit, the lack of disc self-gravity can only be remedied
by reducing the halo density, for example by decreasing $c$.  Thus, models
with a massive thick disc require a smaller value of $c$ than a similar model
with just a thin disc. An enhanced baryon contribution to the gravitational
field in the inner galaxy is also required for satisfaction of the
microlensing constraints, as discussed in Section 4.4 of Paper 3.

Paper 1 showed that lowering $c$ makes discs more unstable and by
$c=4$ the disc is too radially hot.  From Figure 7 of Paper 3, we know that
$c=6-7.5$ gives models which agree with the locally measured DM density.  As
representatives of models with $c<9$ and a thick-IC disc, we here discuss (i)
Q1$\zeta$-, which has $c=6.5$, a more massive IC disc, $M_{\rm disc,
i}=2.5\times10^{10}\msun$, than a P model, and thus a higher final galaxy mass
$M_{\rm f}=6.0\times10^{10}\msun$, and (ii) U1 which has $c=7.5$, 
$M_{\rm disc, i}=2.0\times10^{10}\msun$ and $M_{\rm f}=6.0\times10^{10}\msun$.

Lowering $c$ and increasing $M_{\rm f}$ produces stronger non-axisymmetric
structures. For model Q1 (not shown, see Paper 3), this results in an unrealistically long
bar -- it extends to $R\sim8\kpc$. Lowering $\zeta$ for model Q1$\zeta$- and
thus increasing the number of GMCs yields a hotter thin disc and a weak,
short $R\sim4\kpc$ bar with a strong two arm spiral pattern at outer radii.
The azimuthally averaged and error-corrected radial AVR (black line) is
slightly too hot at most ages on account of the strong spiral pattern, but
the vertical AVR is reasonable for young ages on account of the additional
GMCs compared to the P models studied above.  As in the P models, at old ages
the vertical dispersions are excessive. Interestingly, choosing a Snhd-like
location relative to the bar (blue lines) yields lower dispersions for both
radial and vertical directions. We note that the azimuthal variation is
dominated by the spiral pattern and the location of the Snhd relative to the
bar is likely not relevant here.

The bar of Model U1 extends to $R\sim 6\kpc$ with a very extended X structure
(Figure 12 of Paper 3). Apart from the unrealistically hot vertical dispersions at
old ages caused by the thick disc, its azimuthally averaged AVRs agree well
with Snhd data. The higher vertical dispersions than in other models with
$\zeta=0.08$ are caused by the long bar, as was discussed for the similarly
long and X-shaped bar of Model E2 of Paper 2. The blue lines representing
Snhd-like locations tend to show slightly higher dispersions compared to the
average. So models with thick-disc ICs  with $c\sim7$ indeed provide better
agreement with Snhd data, but their morphologies are not appropriate.

The V models shown in Figure \ref{agevel} live in a $c=6.5$ dark halo and
start from low-mass elliptical ICs, which, however, have negligible impact on
their evolution.  They all have final masses $M_{\rm f}=6\times10^{10}\msun$.
These models fit the Snhd vertical AVR for stars with ages $\tau<7\gyr$
(which are not affected by the thick-disc excess) better than any P
model or the standard Y models of Paper 2.  In these models the shape of
thick-disc excess in both the intrinsic and the corrected AVRs differs from
the corresponding excesses in models with thick ICs because the thick disc
has an intrinsic AVR only in models with declining $\sigma_0$. Also in these
models, input dispersions $\sigma_0$ are at all times higher than the $6\kms$
used in all the four thick-IC disc models shown in Figure \ref{agevel} (see
curves in Figure \ref{sig0}). At early times, high $\sigma_0$ is responsible
for the thick-disc formation, whereas GMC heating determines the final
$\sigma_z$ for thin-disc stars (see Section \ref{sec:thinheat}).

\begin{figure}
\vspace{-0.5cm}
\includegraphics[width=7cm]{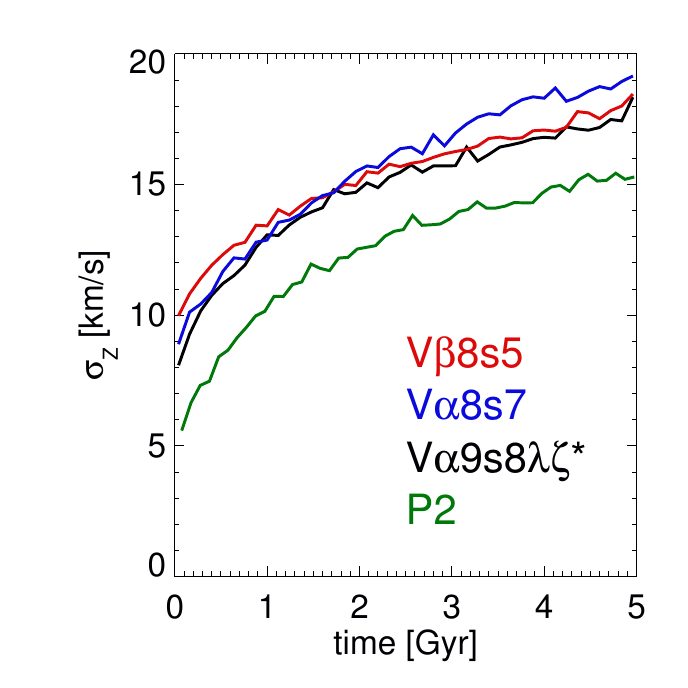}\\
\vspace{-0.5cm}
\caption
{Heating histories $\sigma_z(t)$ for stellar populations which at $t=t_{\rm f}$ live at $R=8\kpc$ and 
are $\tau=5\gyr$ old as traced from their birth to $t=t_{\rm f}$.}
\label{5heat}
\end{figure}

Model V$\beta$8s5 provides a very good match to the radial AVR of the Snhd.
It grows inside out from $h_R=1.5$ to $4.3\kpc$ as $h_R\propto t^{0.5}$ and
has an $\rm{SFR}\propto \exp(-t/8\gyr)$. At $t=t_{\rm f}$ it has a bar to
$R\sim 4\kpc$ and it shows an appropriate level of radial migration over the
last $5\gyr$. Model V$\alpha$8s7 has the same radial growth history $h_R(t)$,
but a different shape of $\sigma_0(t)$ and an ${\rm
SFR}\propto\exp(-t/12\gyr-0.5\gyr/t)$, so the ${\rm SFR}$ grows initially
before peaking at $t=2.5\gyr$ and then declining. Consequently, $\sigma_0$
has to stay high longer to allow for enough thick-disc stars to form. This
explains why the thick-disc excess in $\sigma_z(\tau)$ is shifted to lower
ages. This ${\rm SFR}$ also provides more GMCs and a higher GMC mass fraction
at late times, which is why it has slightly higher $\sigma_z(\tau)$ at low
ages than V$\beta$8s5. In it $\sigma_R$ is slightly too large at most ages,
which is likely connected to its unrealistically large surface density
$\Sigma_b(R=8\kpc)$, similar to that of V$\alpha$9s8$\lambda\zeta$* shown in
Figure 8 of Paper 3, as it otherwise has a bar of reasonable length,
$R\la 4\kpc$, and a reasonable level of radial migration.

Model \V6 grows inside out from $h_R=1.0$ to $3.5\kpc$ as $h_R\propto
t^{0.6}$ and has $\rm{SFR}\propto \exp(-t/6\gyr)$. Both these characteristics
make it more compact than the other V models, especially at early times.  To
bring the GMC numbers at late times, and thus vertical heating, to similar
levels, it has a lower value of $\zeta=0.06$.  It also has an unrealistically high
surface density $\Sigma_b(R=8\kpc)$ and values of $\sigma_R$ that are slightly
too high at intermediate ages. For the oldest ages, it shows the highest
radial dispersions. This comes from the fact that it has $\lambda=1.25$, i.e.
$\sigma_R=1.25\sigma_0$ and thus higher radial input dispersions at early
times. Consequently, the oldest stars in \V6 have unrealistically high values
of both $\sigma_z$ {\it and} $\sigma_R$.

\subsubsection{Thin-disc heating in models with declining $\sigma_0$}
\label{sec:thinheat}

Whereas in models with declining $\sigma_0$ the final velocity dispersions of
old stars are mainly determined by the value of $\sigma_0$ at early times,
the value of $\sigma_z(\tau)$ measured for the thin disc at $t=t_{\rm f}$ and
$R=8\kpc$ is insensitive to $\sigma_0$. Figure \ref{5heat} shows that
thin-disc stars are significantly heated by GMCs. To make this plot we
constructed for
several models heating histories of the stars with ages $\tau=5\pm0.05\gyr$
that at $t_{\rm f}$ are at $R=8\pm0.5\kpc$. As detailed in Paper 2, we
tracked
these stars back in time through the simulation snapshots until their
birth, and at each time determined the velocity dispersion of the population.

According to Figure \ref{sig0}, in the three V models shown in Figure
\ref{5heat} the input dispersion is $\sigma_0=10-15\kms$, but when $\sigma_z$
is measured for a recently born population we obtain a lower value than
$\sigma_0$ as all stars are born at $z=0$ and they will lose vertical kinetic
energy as they all move away from the plane.  Consequently, Figure
\ref{5heat} shows the vertical birth dispersion of these populations to be
$\sigma_z(t_{\rm birth})=8-10\kms$ and thus only slightly higher than that of
young stars in the Snhd today.  As a comparison, the green curve shows
$\sigma_z(\tau)$ in Model P2, in which $\sigma_0=6\kms$. Already Paper 2
showed that slightly increasing $\sigma_0$ improves the fit to the
vertical AVR of the Snhd.

The velocity dispersions in all models shown in Figure \ref{5heat} increase
significantly over the $5\gyr$ from birth to $t=t_{\rm f}$, and in all models
the shape of $\sigma_z(t)$ is similar. The fact that $\sigma_z(t_{\rm
birth})$ is lower in V$\alpha$8s7 than in V$\beta$8s5, but $\sigma_z(t_{\rm
f})$ is higher shows that the details of GMC heating differ from model to
model, as expected.

\begin{figure*}
\hspace{-0.4cm}\includegraphics[width=18cm]{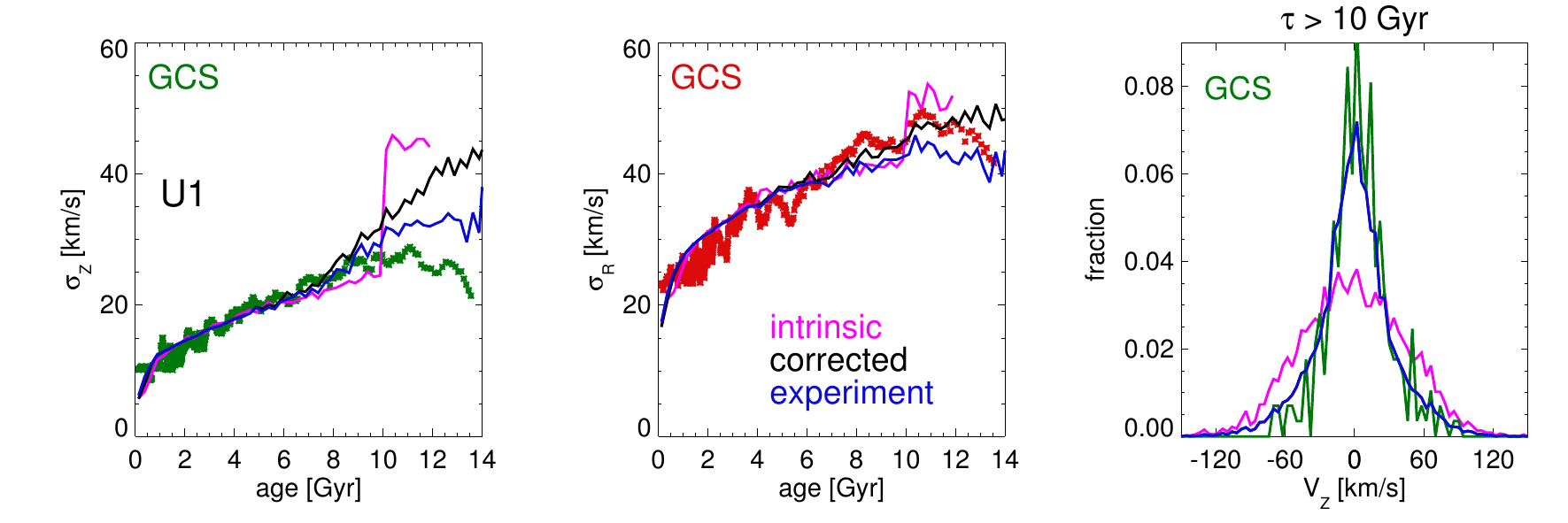}\\
\vspace{-0.4cm}
 \caption {Assigning cool stars to old ages. Left panel: green points are the
GCS vertical AVR $\sigma_z(\tau)$; the pink curve is the intrinsic AVR of
model U1; the black line is the corrected AVR for all azimuths as in Figure
\ref{agevel} and the blue line is the experiment with randomly assigning ages
to 3 per cent of stars from the weighted age distribution.  Middle  panel: the same
for radial AVRs $\sigma_R(\tau)$ with GCS data in red. Right: $V_z$
histograms for stars with $\tau>10\gyr$ in our GCS sample (green),
for true ages from U1 (pink) and for the experiment (blue).} \label{U1exp}
\end{figure*}

\subsubsection{The connection between AVRs and RAVE-TGAS data}

We have seen that Model P2 provides the better representation of the Snhd AVR
for $\sigma_R$, whereas Model \V6 provides the better fit to the AVR for
$\sigma_z$ (Figure \ref{agevel}).  In light of this result, it is interesting
to review the fits these models provide to the velocity distributions of
RAVE-TGAS stars (Figures \ref{hv_P2} and \ref{hv_V6}).  Model P2 provides
reasonable fits to the $V_R$ and $V_z$ distributions when the SFn is ignored
despite Figure \ref{agevel} indicating that its thin disc is too cold
vertically. By contrast, Model \V6 provides good fits to both $V_R$ and $V_z$ when the SFn
is taken into account despite Figure \ref{agevel} indicating that its $V_R$
distributions should be too broad.

These discrepancies might point to a need for a more sophisticated SFn. They
are also possibly connected to the GCS sample (on which Figure \ref{agevel}
depends) being limited to distances $s\la 100\pc$ and having a complicated
SFn in metallicity, which is not modelled here. A simple way out of the
conundrum is to hypothesise that the GCS is an inappropriate indicator for
velocity dispersions at old ages and that our procedure to correct for age
bias and errors underestimates the effects of young stars being classified as
old. In this case, \V6 would be an appropriate model and the SFn applied here
is a reasonable choice given that: (i) \V6 has fitted scaleheights of $h_{z,
{\rm thin}}=284\pc$ and $h_{z, {\rm thick}}=989\pc$ (see Figure \ref{verts})
that are consistent with the MW's vertical profile, (ii) its AVRs for young
stars are in reasonable agreement with the GCS data, and (iii) the SFn
adjusted velocity histograms agree reasonably well at all $|z|$ with the
RAVE-TGAS data.

\subsubsection{Thick-disc stars and the vertical AVR}

The velocity dispersion of the chemically defined thick disc can be as high
as $\sigma_z\approx50\kms$, the specific value depending on $\afe$ and $\feh$
(e.g. \citealp{bovyV}).  In all the models shown in Figure \ref{agevel} the
intrinsic vertical dispersion of the oldest stars is $\sigma_z\sim40-50\kms$,
so consistent with this observation. In models with declining $\sigma_0$, the
specific value depends on the input $\sigma_0(t)$.  In models with a thick-IC
disc, the final velocity dispersion of the oldest stars is largely determined
by the scale height $z_{0,{\rm disc}}\sim1.7\kpc$ of the IC disc, the initial
DM density and the IC baryonic surface density. A comparison of Models P2
and P1s6 also shows that the thick disc of P2 is slightly hotter vertically
than that of P1s6 despite identical ICs. This difference is caused by the
higher contribution of outwards migrators in P2 (Figure \ref{migall}).

Figure \ref{agevel} shows that the models in which old stars ($\tau>10\gyr$)
have the lowest values of $\sigma_z$ are Models M$\beta$1s5 and V$\beta$8s5.
This is connected to the shape of $\sigma_0(t)$ for these models, which drops
more steeply at early times than in the V$\alpha$ models (Figure \ref{sig0}).
According to Figure \ref{verts}, the final vertical profiles of these models
have $h_{z, {\rm thin}}= 270\pc$ and $277\pc$ and $h_{z, {\rm thick}}=
1155\pc$ and $981\pc$, which compare reasonably with the values in the MW.
However, their density ratios $f=0.025$ and $0.038$ and surface density
ratios $f_{\Sigma}=0.11$ and $0.13$ are at the lower end for our models and
problematic if the \citet{juric} values ($f=0.12$, $f_{\Sigma}=0.36$) are
accepted (but see \citealp{bland} for an overview of measurements of these 
ratios for the MW, some of which these models agree with).
Given that the AVR is not sensitive to the intrinsic age distribution
of stars, these ratios could be increased towards the \citet{juric} values by
shifting star formation towards earlier times.

The GCS data do not show such high dispersions for the oldest stars. One
reason will be the specific criteria for the exclusion of halo stars, as was,
e.g., discussed in \citet{casagrande11}. Another possible reason is the
presence of vertically cooler stars at apparent old ages due to (a) seriously
underestimated age errors, and/or (b) the existence of more cold and truly
old stars than predicted by our models, i.e. the velocity distribution of the 
truly old stars being more peaky than modelled. We can test possibilities 
(a) and (b) by adding to a final snapshot cooler stars at the oldest ages.

We thus perform the following simple experiment, which we stress is not
motivated by knowledge of the errors in GCS ages.  As before, we weight model
stars so the weighted true age distribution in the volume $R=8.3\pm0.5\kpc$
and $|z|<100\pc$ agrees with the one present in our GCS sample. For 97 per
cent of the stars from the adjusted age distribution we, as before, assume
age errors of $0.2\tau$ at age $\tau$, but for the remaining 3 per cent we
assume that their measured ages have no information content, so we assign
ages uniformly  redistributed in $\tau \in (0,14) \gyr$.  We apply this
procedure to model U1, as it has a well defined thick disc from its IC and
its thin-disc stars agree reasonably with the Snhd AVRs. It is also among the
models with the largest difference in $\sigma_z(\tau)$ between data and model at
old ages. In this particular model, our procedure implies that of
the stars with $\tau_{\rm assigned}>10\gyr$ only $\sim 45$ per cent are truly
old. We note that the numbers here are adjusted to model U1 and would be 
different for other models. We also note that, according to Figure \ref{agevel}, 
there are models for which the thick disc excess is less severe than in U1.

The results of the experiment are shown by the blue curves in Figure
\ref{U1exp}. In the left panel for $\sigma_z(\tau)$ the upturn at the oldest
ages is significantly less strong in our experiment than with the standard
correction for errors (black curve), also shown in Figure \ref{agevel}. In
fact the difference between the blue curve and the green curve for the GCS
data can be considered a minor issue caused by a slightly inappropriate model,
especially as the GCS downturn at the oldest ages contains only one independent
data point (only every 20th point shown is independent).  Note that the
experiment has no influence at ages $\tau\la10\gyr$ and thus changes no
conclusions about the thin disc. Our experiment makes the agreement at old
ages between the model and the GCS values for $\sigma_R(\tau)$ slightly
worse, but that might be connected to suppressed structure in the very early
stages of adding stars to the thick-IC disc as discussed in connection with Models P2
and P1s6.

The rightmost panel of Figure \ref{U1exp} shows the $V_z$ distributions of
stars with ages $\tau>10\gyr$ for the true ages (pink), our experimentally 
adjusted ages (blue), and the GCS data (green). The blue distribution is 
{\it much} narrower than the pink one as the dispersion at $\tau>10\gyr$ has
been significantly reduced. The green curve from the GCS agrees fairly well 
with the blue experimental curve. This shows that the presence of a vertically 
cooler population of stars in the GCS at old ages in addition to true 
thick-disc stars is plausible. Note that the experimental histogram is 
narrower not only by cooler young stars being identified as old, but by hot 
old stars being classified as young, so removing them from the sample.

Is the offset in $\sigma_z(\tau)$ at $\tau>10\gyr$ between our models and the
GCS data caused by a small fraction of severe unaccounted age mis-determinations or
by the velocity distribution of truly old stars containing more cold stars than in 
our models? This question cannot be answered here but in defence of the GCS ages
we note that \citet{haywood} proposed that there is a population of old, cool 
stars like those that form the core of a more peaky velocity distribution. Because 
of the findings of Paper 1 that vertical disc heating by GMCs and non-axisymmetric 
structures is incapable of scattering stars to the vertical dispersions $\sigma_z\sim40-50\kms$ 
associated with the thick disc, such cool stars would remain vertically cool throughout
the evolution of the disc. The existence of such stars would, however, also require
a reduction of the number of younger thin-disc stars to keep the vertical density profile 
unchanged. On the other hand, \citet{bovyV} showed that each mono-abundance subset 
of stars in SEGUE has a velocity dispersion that is independent of $z$. This finding
is inconsistent with a peaky velocity distribution for mono-abundance subsets,
but the Snhd stars with $\tau>10\gyr$ likely contain a variety of mono-abundance subsets
so that the combined distribution could indeed be peaky.

\section{Discussion}
\label{sec:discuss}

Obviously, our models have shortcomings. As was discussed in Paper 3, they
lack realistic gas components and external heating mechanisms such as
satellite interactions or misaligned infall. Moreover, thick-disc stars are
created ad-hoc and thus correlations between the vertical heating and the
in-plane heating and migration could be missing. Still, from Paper 1 we know
that the thick disc must have been heated prior to formation of the thin
disc. Moreover, our models provide a good representation of both in-plane and
vertical kinematics and provide an appropriate representation for migration
and heating during the thin-disc phase. Most importantly, we are not aware of
any simulations of growing discs which give a better comparison to the
variety of MW data that was discussed here and in Paper 3.  Additionally,
our idealisations give us the opportunity to test a variety of scenarios. Our
set of simulations is thus highly relevant for the study of dynamical
processes which have shaped the MW.

Paper 3 already concluded that models in which the inner few kiloparsecs are
now baryon dominated while baryons and DM contribute equally to the circular
speed $v_{\rm circ}(R_0)$ at the Solar radius, typically include bars similar
to that of the MW bar in terms of length and structure.  Here we have
strengthened this conclusion by showing that these models also fulfil
constraints on the radial AVR of the Snhd from GCS kinematics and on
migration to the Snhd over the last $5\gyr$ from the age-metallicity
relation. Data on the MW circular speed curve $v_{\rm circ}(R)$ from
microlensing measurements towards the Galactic centre \citep{wegg16, cole}
also favour these models. 

The present models require higher baryon-to-DM fractions than the
thin-disc-only models of Paper 1 because the thick disc is massive and its
large velocity dispersions suppress non-axisymmetries. Both radial disc
heating and radial migration are caused by non-axisymmetries, so the
kinematics and extent of radial migration in a model's disc will agree with
data only if the model's vertical and radial distributions of mass also agree
with data. It follows that disc kinematics, chemical heterogeneity and mass
profiles need to be jointly modelled. Moreover, as non-axisymmetries are
continuously excited by accretion onto a galaxy \citep{sellwood84}, it is
essential that such models capture the growth of the disc(s) from early
times. 

\subsection{Radial migration and the chemically defined thick disc}
\label{sec:di:mi}

Analytical models of disc galaxies that combine chemical evolution and
dynamical processes favour a formation scenario for the MW in which the high
$\afe$ Snhd thick-disc stars have migrated outwards from the central Galaxy
and the MW has undergone inside-out formation between the thick-disc
formation stages and now \citep{sb09b, ralph2017}. They have been challenged
by the finding of \citet{vera14,vera16} that thick-disc stars migrate less
than thin-disc stars. \citet{solway} also found that thick-disc stars migrate
less, but concluded that the strength of their radial migration is sufficient
to explain their inner disc origin. The problem with the work of
\citet{vera14} and \citet{vera16} is that neither appropriately considers the
growth of galaxies over cosmological timescales. As a star's age and
chemistry only constrain where it was born, the relevant quantity is the
cumulative angular momentum change since birth and the amount of migration
caused by a specific spiral pattern over a limited time is of minor
importance.

We have therefore studied as a function of age $\tau$ the mean change $\Delta L_z$
in angular momentum experienced by stars that now reside near the Sun -- 
\citet{loebman} determined this for one simulation. For thick-disc stars we find
$\Delta L_z$  depends on the radial growth history of the galaxy. In a model that
starts with a thick-disc scalelength $h_R=2.5\kpc$ and has the same value for the 
input scalelength at all times, we find $\Delta L_z\sim 300\kpc\kms$. In a model 
that grows inside out from $h_{R, {\rm i}}=1.0\kpc$ to $h_{R, {\rm f}}=3.5\kpc$, the
oldest and thickest stellar populations have $\Delta L_z\sim 1300 \kpc\kms$. Hence
in both cases solar-neighbourhood thick-disc stars have moved outwards, and this 
effect is four times larger in a model that grows inside-out. $\Delta L_z$ increases
continuously with increasing age, as was also found by \citet{loebman} (see also 
\citealp{brook12}). We note that \citet{grand} for their hydrodynamical cosmological
simulations of disc galaxy formation have also analysed $\Delta L_z$ vs. $\tau$, but
unfortunately throw all stars irrespective of their final radius in one bowl and 
thus fail to inform us about Snhd-like radii.

Given that chemical evolution models favour inside-out growth, we conclude that 
thick-disc stars with high $\afe$ in the Snhd have likely migrated outwards since
their birth. Another finding points to the same conclusion: if we plot initial angular
momenta $L_{z, {\rm initial}}$ of thick disc stars against present day angular momenta 
$L_{z, {\rm final}}$, we find that there is a maximum $L_{z, {\rm final, edge}}$ that can be 
attained by stars born at low $L_{z, {\rm initial}}$. In our models we find that the values 
of $L_{z, {\rm final, edge}}$ correspond to guiding radii $R_{\rm g}\sim 9-15\kpc$ with the 
highest values found for unrealistically long bars. This finding could explain the
fading of the MW's population with high $\afe$ outside $R\sim11\kpc$ found by
\citet{hayden}.

A related question that has received a lot of attention in the literature is whether
the old outwards migrators are hotter than the inwards- and non-migrators of the 
same age. \citet{sb09b} and \citet{roskar} have advocated a thickening effect of 
outwards migration, whereas \citet{minchev12}, \citet{vera14} and \citet{grand} have
argued against it. First, we should note that in an inside-out forming model, in 
which the oldest populations in the Snhd are dominated by outwards-migrators, this
question is of minor importance. 

In our models, the answer to this question depends strongly on the radial gradient 
of the vertical velocity dispersion $\sigma_z(R)$ in a population of given
age. If $\sigma_z(R)$ is flat, as it is by design in our models with declining 
$\sigma_0$, outwards-migrators are colder vertically than inwards-migrators, as was 
also found by \citet{grand}. The reason is that stars migrating outwards to lower 
surface densities and shallower potential wells cool as a consequence of adiabatic 
conservation of vertical action, and vice versa for inwards migrators 
\citep{sb12,roskar}. However, in thick discs set up with radially constant 
scaleheights $h_z(R)$ and thus declining $\sigma_z(R)$, and in thin discs that are 
heated vertically by GMCs and achieve similarly constant $h_z(R)$, 
outwards-migrators are hotter than non- and inwards-migrators because adiabatic 
cooling during outwards-migration is insufficient to cancel the gradient in 
$\sigma_z(R)$. The question about whether outwards migration thickens the disc is thus
not a question of migration but of the vertical heating mechanism.

An associated question is the origin of the observed negative vertical
metallicity gradient in the Snhd, $d\feh / dz \approx -0.25 \;\rm{dex}/\kpc$ \citep{schlesinger, hayden14}.
The stars in our models do not carry metallicity information. However, in Paper 3 we showed that
the vertical age gradient in the Snhd for inside-out growing models with declining $\sigma_0$
is in agreement with recent measurements of \citet{casagrande16}. In the inside-out growing 
chemodynamical evolution models of \citet{ralph2017}, the Snhd today contains a large number of old 
and kinematically hot, thick-disc stars which originate from the inner galaxy, in agreement with 
our models. In the Snhds of these models, thick-disc stars have high $\afe$ and lower $\feh$ than 
young thin-disc stars, so the vertical $\feh$ gradients agree with observations
(see also \citealp{kawata17}). It is thus reasonable to assume that our models are not 
in disagreement with these observations. It would be interesting to combine our models with a 
prescription for chemical evolution in a future paper.

\subsection{Disc heating and Snhd kinematics}
\label{sec:di:ki}

We have confirmed the conclusion of Papers 1 and 2 that the vertical heating
of the thin disc is well explained by scattering of stars off GMCs, as
originally envisioned by \citet{spitzer}. The presence of a thick disc
slightly weakens the effect of GMCs, mainly because less star formation
occurs in the thin-disc phase when there is a thick disc, so there are fewer
GMCs. However, it was already noted in Papers 1 and 2 that our standard value
for the SF efficiency $\zeta=0.08$ and the input velocity dispersion
$\sigma_0$ yield lower vertical scaleheights $h_{z, {\rm thin}}$ than those
observed in the MW. The same is true for models with thick discs, but, as
discussed in Paper 2, slightly lowering $\zeta$ and/or slightly increasing
$\sigma_0$ does not violate observational constraints and yields better
agreement with the vertical AVR and scaleheights. We note that
\citet{gustafsson} have recently confirmed our conclusions regarding vertical
thin disc heating by GMCs.

A remaining problem for the vertical AVR $\sigma_z(\tau)$ is the lack of a
thick-disc signal in the GCS data. When the survey's bias to younger stars
and errors are taken into account, the intrinsic discontinuity in $\sigma_z$
around the age corresponding to the onset of thin-disc formation is weakened
(see also \citealp{martig14}) but the predicted relation never becomes as
flat with age as that seen in the data.  We have shown that this problem can
be alleviated by assuming that a few per cent of stars have been assigned
seriously erroneous ages. This hypothesis can bring models and observations
into agreement, because if even a small fraction of the large numbers of
young stars in the GCS are scattered to old ages, the scattered stars
comprise a significant fraction of all apparently old stars. An alternative
explanation is that the Snhd contains more truly old and vertically cold stars 
than is predicted by our models, in which all old stars are part of a thick
disc with a broader vertical velocity distribution. Moreover, uncertainties
in the exclusion of halo stars from the GCS sample and metallicity dependent
survey selection effects could play a role.

As the data from the GCS are spatially very limited, we have also compared
our models to velocity distributions from RAVE-TGAS. At radii $R\sim R_0$
our model histograms for $V_R$ and $V_\phi$ vary significantly with
azimuth. By shifting each histogram horizontally to optimise the fit to the
corresponding histogram of Galactocentric velocities, we obtain estimates of
the solar motion with respect to the Galactic centre. On account of the bar
and spiral structure these estimates vary from azimuth to azimuth by up to
$\sim\pm10\kms$ (see also the models of \citealp{monari}).  It is thus
important to choose an appropriate azimuthal location and avoid azimuthal
averaging when comparing simulations to Snhd data. As knowledge on spiral
structure is limited, we merely choose locations that are positioned relative
to the bar like the Snhd. 

A significant complication when comparing the models to data is that one
should take into account the SFn of the relevant survey, which will depend on
age, metallicity and location. We have used a simple approximation to the age
dependence and minimised the impact of other dependencies by slicing samples
in vertical bins. Age selection is particularly important at intermediate
altitudes $|z|\sim 200-500\pc$ because it favours younger and thus vertically
colder stars, and at these altitudes significant numbers of both old and
young stars are present. Generally models that are favoured by structural and
AVR constraints, compare well with RAVE-TGAS data, but if the age SFn is
taken into account, the RAVE-TGAS data favour slightly hotter models
(see also \citealp{SandersB15}). As the volumes probed by RAVE-TGAS and GCS
differ and for each survey we consider a different type of data (AVR vs.
velocity distributions), it is not surprising that the models that provide
the best fits to the GCS do not necessarily provide the best fits to
RAVE-TGAS.

\section{Conclusions}
\label{sec:conclude}

We have analysed a set of idealised $N$-body simulations of growing disc
galaxies with thin and thick discs to gain a better understanding of how disc
heating and radial migration shape a present-day disc galaxy.  Thick discs in
these models are either represented by thick-IC discs or created by adding
stars with continuously declining birth velocity dispersions $\sigma_0(t)$.
It was shown in Paper 3 that both types of models produce galaxies with 
double-exponential vertical density profiles. 

Our main conclusions regarding radial migration are:
\begin{itemize}

\item{Models with an appropriate amount of non-axisymmetric structures provide
just the right level of radial migration needed to explain the level of chemical
diversity in the Snhd.}

\item{In appropriate models, thick disc stars at $R\sim R_0$ and $t=t_f$ have
on average gained angular momentum $L_z$ since their birth. The typical
amount of angular momentum gain $\Delta L_z$ depends on the model's
radial-growth history, being larger for inside-out formation than for
models with constant radial birth scalelength.}

\item{Whether old outwards migrators heat the disc vertically at $R\sim R_0$
depends on $\sigma_z(R)$, the radial profile of the vertical velocity
dispersion of thick disc stars. If $\sigma_z(R)$ is constant, outwards
migrators will be vertically cooler than non-migrators, if the vertical
scaleheight $h_z$ does not vary with $R$ and thus $\sigma_z$ declines with
$R$, outwards migrators will be hotter than non-migrators.}

\item{There is an effective upper boundary in angular momentum 
$L_{z, {\rm final, edge}}$ that thick-disc stars originating in the inner galaxy can 
reach by radial migration over their lifetimes. It typically corresponds to 
guiding centre radii $R_{\rm g}\sim10-15\kpc$ and can explain the fading of stars of the 
high $\afe$-sequence outside $R_0$.}

\item{The chemically defined thick disc stars in today's Snhd have likely 
migrated outwards from the inner Galaxy over their lifetime.}
\end{itemize}

Radial migration and radial disc heating are both caused by non-axisymmetries,
which is why models favoured by our analyses of radial migration are also 
favoured by data on Snhd kinematics. From our analyses of the latter we find:
\begin{itemize}

\item{Our models show significant azimuthal variation of the $V_R$ and 
$V_{\phi}$ histograms at $R_0$, both in shape and peak velocity. These variations are
caused by the bar and spiral arms. Streaming velocities can vary by as much 
as $\sim \pm 10\kms$.}

\item{For comparisons between models and Snhd data it is essential to take
into account this azimuthal variation and choose a Snhd-like position relative to
the bar (and ideally also the spiral arms). Moreover, survey selection functions 
in age and distance have to be considered.}

\item{Models with appropriate levels of non-axisymmetries and radial migration
provide good fits to velocity histograms from RAVE-TGAS at all altitudes 
$|z|\le 800 \pc$ for which reasonably precise data are available.}

\item{Such appropriate models also provide the right amount of radial disc heating
as traced by the radial AVR $\sigma_R(\tau)$.}

\item{In both models with declining $\sigma_0$ and models with thick-disc ICs, 
the vertical AVRs $\sigma_z(\tau)$ of thin-disc stars are shaped by GMC heating, as
in thin-disc-only models.}

\item{GMCs heat thin-disc stars vertically less effectively in a model that has a
thick disc than in the model that has  the same dark halo and final
disc mass but only a thin disc
because (a) the thick component impedes the development of non-axisymmetries,
and (b) formation of the thin disc requires smaller SFRs and consequently a
lower GMC mass fraction.}

\item{Models with appropriate thick discs yield values of
$\sigma_z(\tau)$ at $\tau>10\gyr$ that exceed those observed in the Snhd.
Possible explanations are either unaccounted age errors scattering young,
cool stars to high ages or a higher number of truly old and cold disc 
stars than predicted by our models.}
\end{itemize}

To achieve agreement between models and data regarding non-axisymmetric structures, radial
migration {\it and} disc heating, models which are baryon dominated in the centre and have 
roughly equal contributions to the circular-speed curve at $R_0$ are favoured. Such models
require galaxy masses $M_{\rm f}\approx5-6\times10^{10}\msun$ and DM halo concentration 
parameters $c\approx7$ for haloes with masses $M_{\rm DM}=10^{12}\msun$. As was already noted in Paper
1, these numbers agree reasonably with what is expected for a MW-like galaxy in a $\Lambda$ 
cold dark matter cosmology.

Here and in Paper 3 we have analysed how our $N$-body models of growing disc
galaxies with thin and thick discs compare to MW data as regards density
profiles, baryonic and DM contributions to the circular-speed curve, the bar,
the age structure, Snhd velocity distributions and AVRs, and radial
migration, which is constrained by the chemistry of stars. We have
deliberately avoided going into details of the dynamical processes that drive
the models but have instead focused on the general ingredients necessary to
construct a realistic model. It has become clear that realistic structural
properties correlate well with realistic levels of disc heating and
migration.  Although we have not been able to single out one model as a
particularly suitable MW-analogue, we have narrowed the range of possible
scenarios. It will be interesting to use these models for more detailed
studies.

\section*{Acknowledgements}
We thank the referee for comments that helped improve the paper.

This work was supported by the UK Science and Technology Facilities Council (STFC)
through grant ST/K00106X/1 and by the European Research Council under the European 
Union's Seventh Framework Programme (FP7/2007-2013)/ERC grant agreement no.~321067.

This work used the following compute clusters of the STFC DiRAC HPC Facility 
(www.dirac.ac.uk): i) The COSMA Data Centric system at Durham University, operated
by the Institute for Computational Cosmology. This equipment was funded by a BIS 
National E-infrastructure capital grant ST/K00042X/1, STFC capital grant 
ST/K00087X/1, DiRAC Operations grant ST/K003267/1 and Durham University. 
ii) The DiRAC Complexity system, operated by the University of Leicester 
IT Services. This equipment is funded by BIS National E-Infrastructure capital 
grant ST/K000373/1 and STFC DiRAC Operations grant ST/K0003259/1.
iii) The Oxford University Berg Cluster jointly funded by STFC, the Large 
Facilities Capital Fund of BIS and the University of Oxford. 
DiRAC is part of the National E-Infrastructure. 

This work has made use of data from the European Space Agency (ESA) mission 
Gaia (https://www.cosmos.esa.int/gaia), processed by the Gaia Data Processing 
and Analysis Consortium (DPAC, https://www.cosmos.esa.int/web/gaia/dpac/consortium). 
Funding for the DPAC has been provided by national institutions, in particular 
the institutions participating in the Gaia Multilateral Agreement.

\appendix

\section{Basic properties of the models}
\label{appendix}

In this appendix, we present several Figures, which show basic properties of all the models studied
in this paper and are helpful for the understanding of our analyses. Further information on details
of the Figures and on the evolution histories of the model galaxies can be found in Papers 1, 2 and 3.

\begin{figure*}
\hspace{-0.cm}\includegraphics[width=18cm]{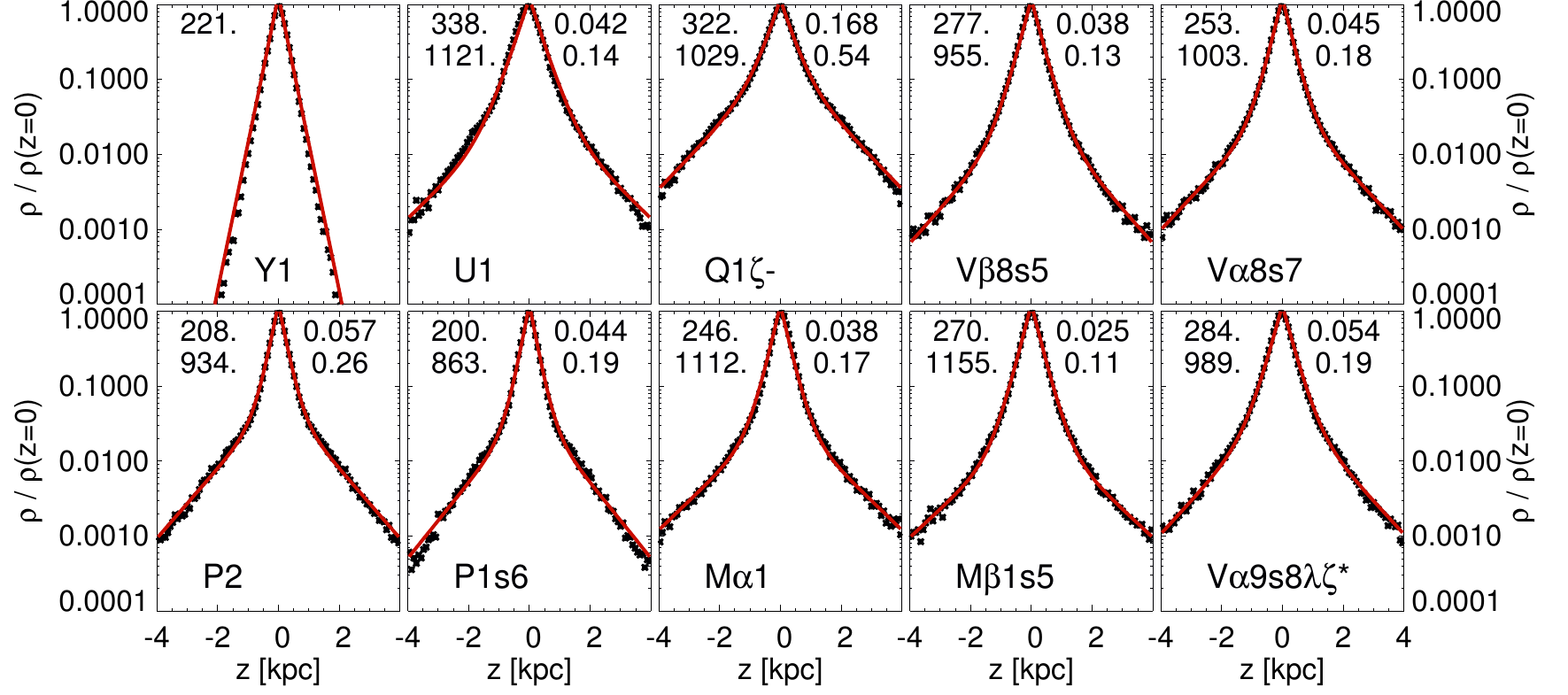}\\
\vspace{-0.1cm}
\caption
{Vertical profiles of models at $t=t_{\rm f}$ and $R=8\pm0.5\kpc$ are shown as black points. 
Overplotted are fits of $\rho(z)=\rho_0[\exp(-|z|/h_{\rm thin})+f\exp(-|z|/h_{\rm thick})]$ to 
these profiles. The numbers in the upper left corners are the values of the thin and thick
disc scaleheights $h_{\rm thin}$ and $h_{\rm thick}$ in pc. The numbers in the upper right corners
are the values of the density ratio $f$ and the surface density ratio 
$f_{\Sigma}=f h_{\rm thick} / h_{\rm thin}$. Y1 has no thick disc.}
\label{verts}
\end{figure*}

\begin{figure*}
\hspace{-0.cm}\includegraphics[width=18cm]{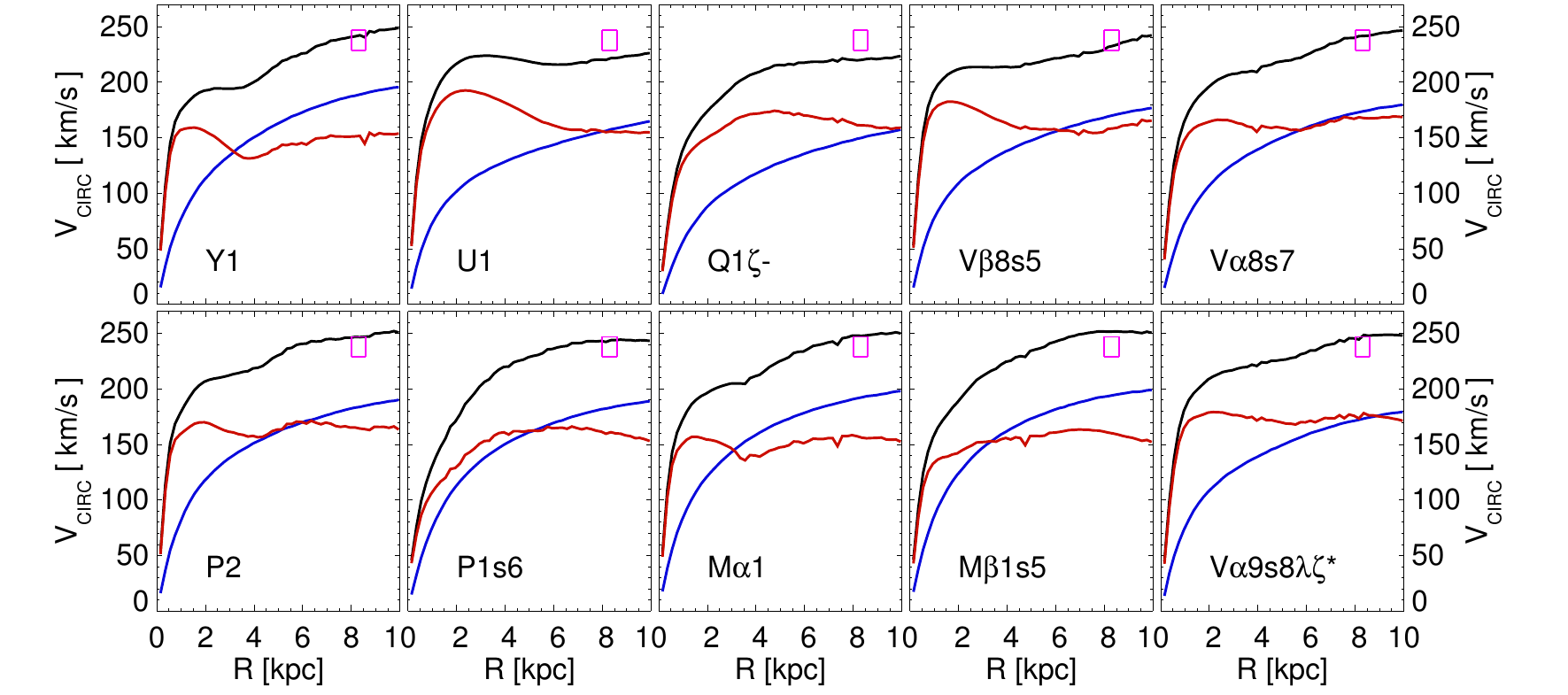}\\
\vspace{-0.1cm}
\caption
{Circular speed curves $v_{\rm circ}(R)$ measured in the midplane of the disc. We determine the circular
speed as $v_{\rm circ}=\sqrt{{a_R(R)} R}$, where $a_R(R)$ is the azimuthal average of the radial 
gravitational acceleration, $\partial\Phi/\partial R$. Blue lines mark the contributions from DM 
and red lines the baryonic contribution. Pink boxes mark the constraints on $R_0$ and 
$v_{\rm circ}(R_0)$ determined by \citet{ralph2012}.}
\label{vcirc}
\end{figure*}

\begin{figure*}
\vspace{-0.cm}
\hspace{-0.cm}\includegraphics[width=18cm]{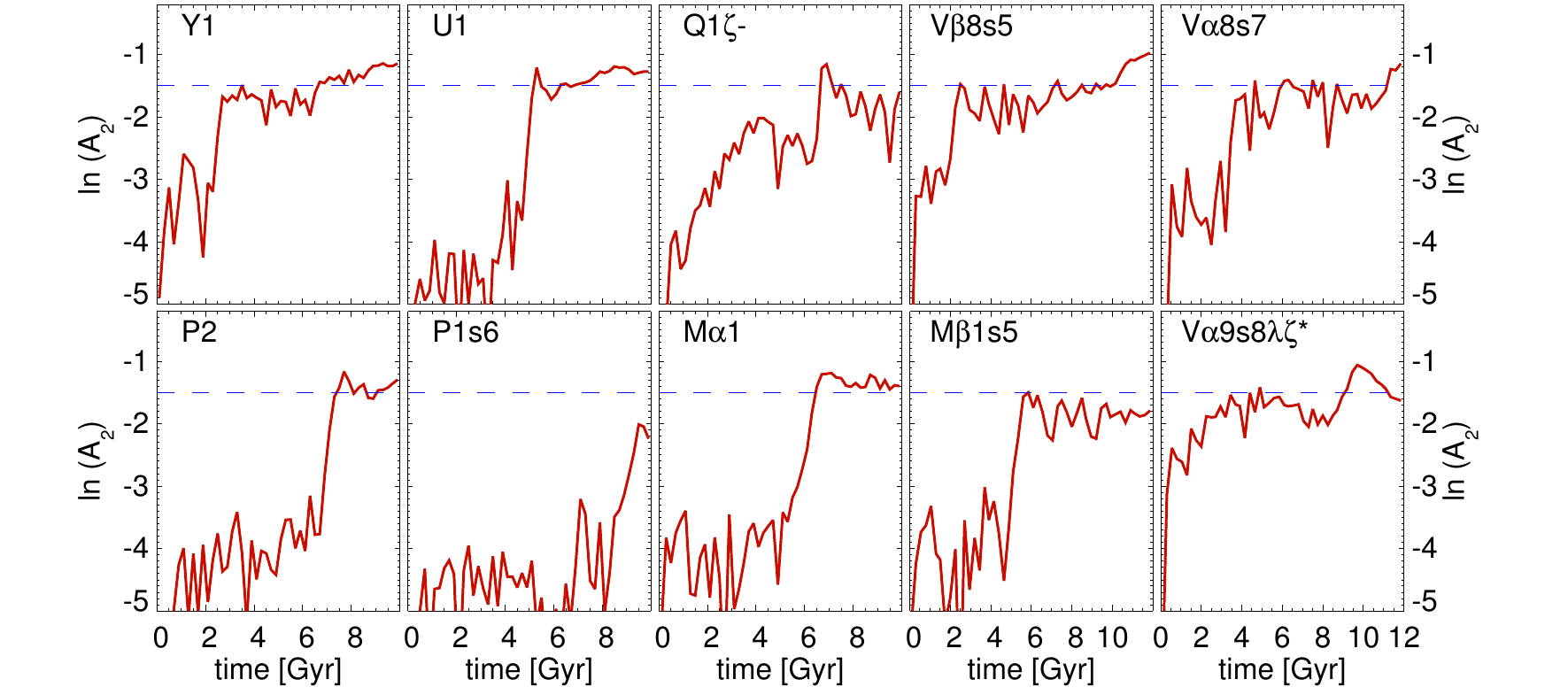}\\
\caption
{Evolution with time of $m=2$ Fourier amplitude $A_2$ measured within $R=3\kpc$. The blue dashed lines mark 
$\ln(A_2)=-1.5$.}
\label{barx}
\end{figure*}

\begin{figure*}
\vspace{-0.cm}
\hspace{-0.cm}\includegraphics[width=18cm]{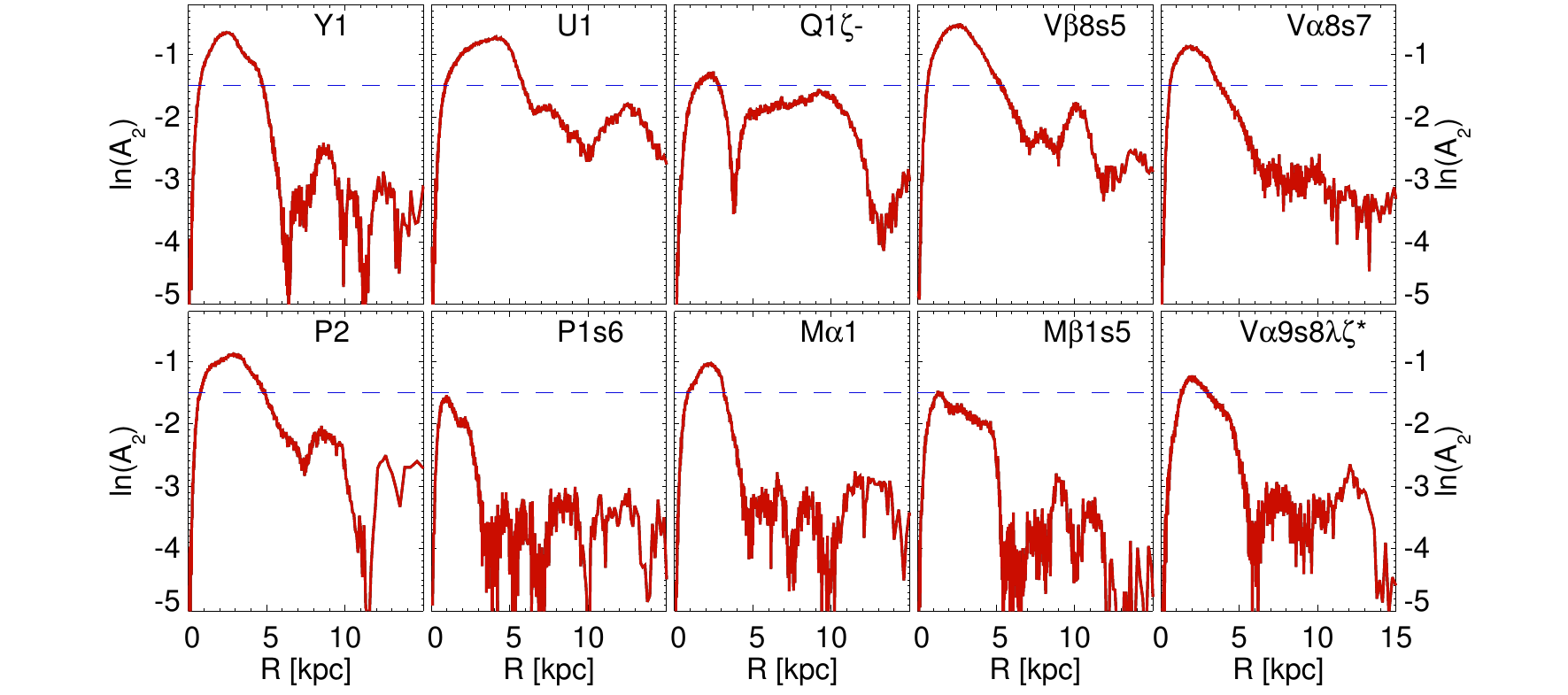}\\
\caption
{Radial profiles of the $m=2$ Fourier amplitude $A_2(R)$ measured at $t=t_{\rm f}$. The blue dashed lines mark 
$\ln(A_2)=-1.5$.}
\label{sd}
\end{figure*}

\label{lastpage}
\end{document}